\documentclass[aps,twocolumn,superscriptaddress]{revtex4}
\usepackage{mathrsfs}
\usepackage{array}
\usepackage{amsmath}
\usepackage{mathtools}
\usepackage{graphicx}
\usepackage{subfigure}
\usepackage{amstext}
\usepackage{amsfonts}
\usepackage{amssymb}
\usepackage{bm}
\usepackage{epstopdf}

\usepackage[usenames,dvipsnames]{color}
\usepackage{hyperref}
\usepackage{calc}
\begin{document}

\title{Exact Master Equation for Quantum Brownian Motion with Generalization to Momentum-Dependent System-Environment
Couplings}
\author{Yu-Wei Huang}
\affiliation{Department of Physics, and Center for Quantum Information Science, National Cheng Kung University, Tainan
70101, Taiwan}
\author{Wei-Min Zhang}
\email{wzhang@mail.ncku.edu.tw}
\affiliation{Department of Physics, and Center for Quantum Information Science, National Cheng Kung University, Tainan
70101, Taiwan}
\affiliation{Physics Division, National Center for Theoretical Sciences, Taipei 10617, Taiwan}

\begin{abstract}
In this paper, we generalize the quantum Brownian motion to include momentum-dependent system-environment couplings. The
resulting Hamiltonian is given by
$
    H_\textsl{tot}
    =
        \hbar \omega_\textsc{s} a^\dagger a
        + \sum_k \hbar \omega_k b_k^\dagger b_k
        + \sum_k \hbar (V_k a^\dagger b_k + V_k^* b_k^\dagger a)
        + \sum_k \hbar (W_k a^\dagger b_k^\dagger + W_k^* b_k a)
$.
The conventional QBM model corresponds to the spacial case $W_k = V_k$. The generalized QBM is more complicated but the
generalization is necessary. This is because the particle transition and the pair production between the system and the
environment represent two very different physical processes, and usually cannot have the same coupling strengths. Thus,
the conventional QBM model, which is well-defined at classical level, is hardly realized in real quantum physical world.
We discuss the physical realizations of the generalized QBM in different physical systems, and derive its exact master
equation for both the initial decoupled states and initial correlated states. The Hu-Paz-Zhang master equation of the
conventional QBM model is reproduced as a special case. We find that the renormalized Brownian particle Hamiltonian
after traced out all the environmental states induced naturally a momentum-dependent potential, which also shows the
necessity of including the momentum-dependent coupling in the QBM Hamiltonian. In the Hu-Paz-Zhang master equation, such
a renormalized potential is misplaced so that the correct renormalization Hamiltonian has not been found. With the exact
master equation for both the initial decoupled and and initial correlated states, the issues about the initial jolt
which is a long-stand problem in the Hu-Paz-Zhang master equation is also re-examined. We find that the so-called
``initial jolt'', which has been thought to be an artificial effect due to the use of the initial decoupled
system-environment states, has nothing do to with the initial decoupled state. The new exact master equation for the
generalized QBM also has the potential applications to photonics quantum computing.
\end{abstract}

\pacs{}
\maketitle

\section{Introduction}
\label{sec:intro}
As it is well-known, any realistic system inevitably interacts with its environment. For nano-scale quantum devices or
more general mesoscopic and quantum systems, such interactions are usually not negligible, and thus these systems must
be treated as open quantum systems. The essential physics of open quantum systems is the entanglement generation between
the system and the environment so that the system cannot be retained in pure quantum states. This makes the system lose
its quantum coherence, known as quantum decoherence which is the key obstacle in the development of quantum technology.
Up to date, theoretical description of open quantum systems is still a big challenge in both fundamental research and
practical applications. The major difficulty for solving the dynamics of open quantum systems is how to find the
physically reasonable system-environment coupling which is usually unknown in priori, and how to derive unambiguously
the equation of motion, called the master equation, for the dynamical evolution of open quantum systems in terms of the
reduced density matrix which is not always computable.

Historically, a prototype of open quantum systems is the quantum Brownian motion (QBM) \cite{Grabert1988,Weiss1992}.
Quantum mechanically, the Brownian particle is modeled as a harmonic oscillator that linearly couples to the environment
which is made of a continuous distribution of infinite number of harmonic oscillators, as originally proposed in the
seminal works of Feynman and Vernon \cite{Feynman1963} and also Caldeira and Laggett \cite{Leggett1983a,Leggett1983b}.
The system-environment coupling in QBM was assumed as a strict linear coupling $\sum_k \! C_k x q_k$, where $x$ and
$q_k$ are the position coordinates of the system and environment oscillators, respectively, and $C_k$ is the coupling
amplitude. The exact master equation of the QBM for such a system-environment coupling has been derived during 1980's
$\sim$ 1990's with different approaches \cite{Leggett1983a,Haake1985,Hu1992,Karrlein1997}. Correspondingly, the
dissipative dynamics of QBM has been extensively investigated in a rather broad physical research areas
\cite{Grabert1988,Weiss1992}. In particular, the classical dissipative dynamics, the fluctuation-dissipation theorem and
the classical Langenvin equation have been deduced from this quantum Brownian motion in the weak-coupling and high
temperature limits \cite{Grabert1988,Weiss1992,Feynman1963,Leggett1983a,Leggett1983b}.

Microscopically, the above classically well-defined linear coupling $\sum_k \! C_k x q_k$ depicts the Brownian particle
dissipation and fluctuations through energy exchange between the system and the environment. However, probably one does
not realize that this system-environment coupling is physically not self-consistent at quantum level. Specifically, the
Hamiltonian of a harmonic oscillator can be written as $H_\textsc{s} = \hbar \omega_\textsc{s} a^\dagger a$, where
$\hbar \omega_\textsc{s}$ is the quanta of the harmonic oscillator energy of the system, and $a^\dagger$ and $a$ are the
corresponding creation and annihilation operators obeying commutation relation $[a, a^\dagger] = 1$. Similarly, the
Hamiltonian of the environment made by infinite number of harmonic oscillators can be expressed as $H_\textsc{e} =
\sum_k \hbar \omega_k b_k^\dagger b_k$, where $b_k^\dagger$, $b_k$ are the creation and annihilation operators of the
quantum oscillator mode $k$ with frequency $\omega_k$. Consequently, the linear system-environment coupling Hamiltonian
can be expressed as
$
    \sum_k \! C_k x q_k
    =
        \sum_k \! \hbar V_k (
            a^\dagger b_k
            + b_k^\dagger a
            + a^\dagger b_k^\dagger
            + b_k a
        )
$
with $V_k = C_k / 2 \sqrt{M m_k \omega_\textsc{s} \omega_k}$, where $M$, $m_k$ are the masses of the system oscillator
and the environmental oscillator mode $k$. It shows that the first two terms, proportional to $a^\dagger b_k +
b_k^\dagger a$, depict the energy exchange (energy transition) between the system and the environment. While the last
two terms, proportional to $a^\dagger b_k^\dagger + b_k a$, describe the processes of generating and annihilating
separately two quanta of energy out of the blue, from nothing. As one knows, the energy quanta of harmonic oscillator
can be realized with photons or phonons, and physically one can generate or annihilate two photons (phonons) through
non-linear crystals or atomic virtual states, but these physical processes are very different from the single photon
transition processes of the first two terms in linear quantum optics. In other words, in reality, the coupling strength
$V_k$ for the single particle transition processes is usually not the same as that of the two particle generation or
annihilation.

Analogy to the above linear coupling between the system and the environment for the QBM, a similar situation also occurs
in the canonical quantization of light-matter interaction. In quantum optics, under the dipole momentum approximation,
the interesting light-matter interaction is mainly determined by $- \bm{d} \cdot \bm{E}(x,t)$, where $\bm{d} = q \bm{r}$
is the atomic dipole moment and $\bm{E}(x,t)$ is the radiative electric field. Treating approximately the atom as a
two-level system, the above light-matter interaction can be written quantum mechanically as
$
    - \bm{d} \cdot \bm{E}(x,t)
    =
        \sum_k g_k \sigma^x (b_k + b_k^\dagger)
    =
        \sum_k g_k (
            \sigma^+ b_k
            + \sigma^- b_k^\dagger
            + \sigma^+ b_k^\dagger
            + \sigma^- b_k
        )
$,
where $\sigma^x = \sigma^- + \sigma^+$ is the Pauli matrix, and $b_k^\dagger$, $b_k$ are the creation and annihilation
operators of photons for the radiative field mode $k$. In this light-matter interaction, the first two terms
proportional to $\sigma^+ b_k + \sigma^- b_k^\dagger$ depict the electron transitions in atoms from the low-energy state
to the high-energy state by absorbing a photon, and from the high-energy state back to the low-energy state by emitting
a photon. These are the fundamental physical processes for light-matter interaction. While, there are other two terms,
proportional to $\sigma^+ b_k^\dagger + \sigma^- b_k$, which describe the electron transitions from the low-energy state
to the high-energy state by emitting a photon, and from the high-energy state to the low-energy state by absorbing a
photon. Actually, the last two processes barely occur due to the law of energy conservation, see Fig.~\ref{fig:trans},
also see more discussions given in the next section.

In the literature, these physically infeasible processes in the linear coupling of the QBM and in the light-matter
interaction of quantum optics are often called the counter-rotating terms. They can usually be eliminated under the
so-called rotating wave approximation. In the high-temperature and weak-coupling limit, it can be shown that the
contributions from these counter-rotating terms are negligible. However, in the strong coupling and low temperature
regime, the counter-rotating terms could make a significant effect. In recent years, many effects have been devoted to
search the counter-rotating effects theoretically and also experimentally
\cite{Zheng2008,Niemczyk2010,Li2013,Garziano2016, Xu2016,Wang2017,Yoshihara2017}. Here we should point out that these
counter-rotating terms either are forbidden by the energy conservation law for every individual microscopic physical
processes, or can be modified by the comprehensive physical description of the system-environment couplings.
Specifically, for an example, for the radiative electromagnetic field, $\bm{E} = - \partial \bm{A} / \partial t$ and
$\bm{B} = \nabla \times \bm{A}$ where the vector potential $\bm{A}$ and the electric field $\bm{E}$ are the generalized
coordinate and momentum in quantum electrodynamics (QED). Besides the atomic dipole moment interaction with the electric
field, there is also the atomic magnetic dipole moment interaction with the magnetic field. Both interactions come from
the position-momentum dependent atom-light interaction $- \frac{q}{m} \bm{A} \cdot \bm{P}_\textsl{e}$, where
$\bm{P}_\textsl{e}$ is the electron momentum \cite{Tannoudji1998}. For the QBM, besides the linear coupling between the
oscillator coordinates of the system and the environment, there are also the momentum-dependent system-environment
couplings. When these contributions are elaborated, the inconsistent processes in system-environment couplings of open
quantum systems can be diminished.

In this paper, we shall provide first a detailed analysis of the system-environment couplings for open quantum systems
in general, and then generalize the QBM by including the momentum-dependent coupling to get rid of the possible
ambiguity of system-environment couplings at quantum level. We shall also discuss practical realizations of such a
generalization. In Sec.~\ref{sec:EME}, we will derive the exact master equation of the generalized QBM for both the
initial decoupled and initial correlated states. We shall also explore the relation of dissipation and fluctuation
dynamics in open quantum systems with the decay and diffusions of the Brownian particle. In Sec.~\ref{sec:QBM}, we will
show how the previous Hu-Paz-Zhang master equation for the QBM without including momentum-dependent coupling
\cite{Leggett1983a,Haake1985,Hu1992,Karrlein1997} can be reproduced as a very special case. We will also re-examine some
long-standing debated issues about QBM, such as the origin of the initial jolt phenomena and the initial decoupled state
problem one often questioned for the study of the QBM with the exact master equation. Finally, discussions and
perspectives are given in Sec.~\ref{sec:conclusion}. In Appendix~\ref{sec:UV}, the nonequilibrium Green functions for
open quantum systems we previously introduced are generalized to open systems containing pairing couplings through the
Heisenberg equation of motion. The generalized nonequilibrium fluctuation-dissipation theorem in the time-domain is also
derived there.


\section{A General Analysis of Momentum-Dependent System-Environment Couplings}
\label{sec:gen_an}
In the early studies of quantum dissipative dynamics, system-environment couplings in open systems are mainly modeled
based on classical mechanics. The prototype example is given in the pioneering work by Feynman and Vernon for developing
their influence functional theory for QBM \cite{Feynman1963}. In the exploration of the dissipative and noise effects of
other interaction systems on the system of interest, Feynman and Vernon proposed a general test system $x$ interacting
to linear systems $\{ q_k \}$ (an oscillator bath) with the Lagrangian
\begin{align}
    L_\text{FV}
    \!\! =
        \! L_\textsc{s}(\dot{x}, x)
        \! + \! \frac{1}{2} \sum_k \! m_k (\dot{q}_k^2 \! - \! \omega_k^2 q_k^2)
        \! - \! U(x) \! \sum_k \! C_k q_k
    .
\label{FVM}
\end{align}
Without loss of generality, one usually takes a linear coupling $U(x) = x$. The oscillator bath is an environment
consisting of a collection of harmonic oscillators with a continuous frequency distribution. Such an environment can be
easily realized as radiative electromagnetic fields or vibrations of lattices in matter. The Lagrangian formulation in
terms of the dynamical variables of positions and velocities is setup in order to conventionally use the Feynman's path
integrals for the trace over all the environmental states \cite{Feynman1965}. In such circumstances, the path integrals
over environmental degrees of freedom can be done exactly. Thus the environment effect on the test system is given by an
influence functional of coordinates of the system only. It is equivalent to external semiclassical dissipative and
fluctuating (random) forces acting on the test system. As Feynman and Vernon pointed out \cite{Feynman1963}, if a
general non-linear interaction system $q_k$ is weakly coupled to a test system $x$ with a more general interaction
potential $U(x) U'(\{ q_k \})$ which is small, then the effect on the test system is that of a sum of oscillators whose
frequencies correspond to the possible transitions of the interaction system. Thus, to the extent that second order
perturbation theory yields sufficient accuracy, the effect of an interaction system is that of a linear system
\cite{Feynman1963}.

In 1980's, Caldeira and Leggett made further investigations to QBM and quantum tunneling of dissipative systems, in
which similar linear couplings are made in order to derive a dissipative (damping) equations of motion from quantum
mechanics \cite{Leggett1983a,Leggett1983b}. Specifically, to consistently produce the semiclassical damped equation of
motion, $M \ddot{x} + \eta(x) \dot{x} + \mathrm{d} V / \mathrm{d} x = F_\textsl{ext}(t)$ where $F_\textsl{ext}(t)$ is an
external force, Caldeira and Leggett also proposed the environment as a continuous frequency distribution of a set of
harmonic oscillators interacting linearly with the system,
\begin{align}
    L_\text{CL}
    ={}
        & \frac{1}{2} M \dot{x}^2
        \! - \! V(x)
        \! + \! \frac{1}{2} \sum_k m_k (\dot{q}_k^2 \! - \! \omega_k^2 q_k^2) \notag \\
        & \! - \! \sum_k F_k(x) q_k
        \! - \! \sum_k F_k^2(x) / 2 m_k \omega_k^2
    .
\label{CLM}
\end{align}
When the system is a tunneling system, Eq.~(\ref{CLM}) is often called the Caldeira--Leggett model in the literature.
Except for the last term which is added as a counter-term for the stability of potential renormalization,
Eq.~(\ref{CLM}) is equivalent to Eq.~(\ref{FVM}). For $x$-independent dissipation $\eta(x) = \eta$, the semiclassical
damped equation of motion can be realized with the strict linear coupling between the system and the environment $F_k(x)
= C_k x$. Other more complicated coupling mechanisms are analyzed extensively in the Appendix C of \cite{Leggett1983b}
but these analysis are indeed superfluous from the microscopical level of interacting systems, as Caldeira and Leggett
pointed out in their own work \cite{Leggett1983b}.

\begin{figure}
\subfigure[~feasible processes]{
    \includegraphics[width=0.45\linewidth]{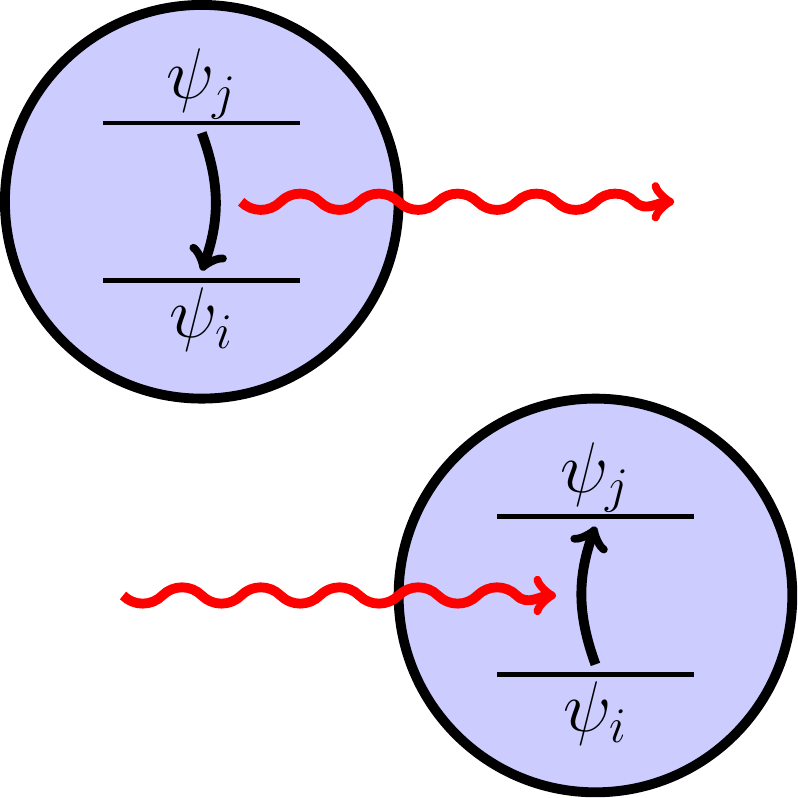}
}
\subfigure[~infeasible processes]{
    \includegraphics[width=0.45\linewidth]{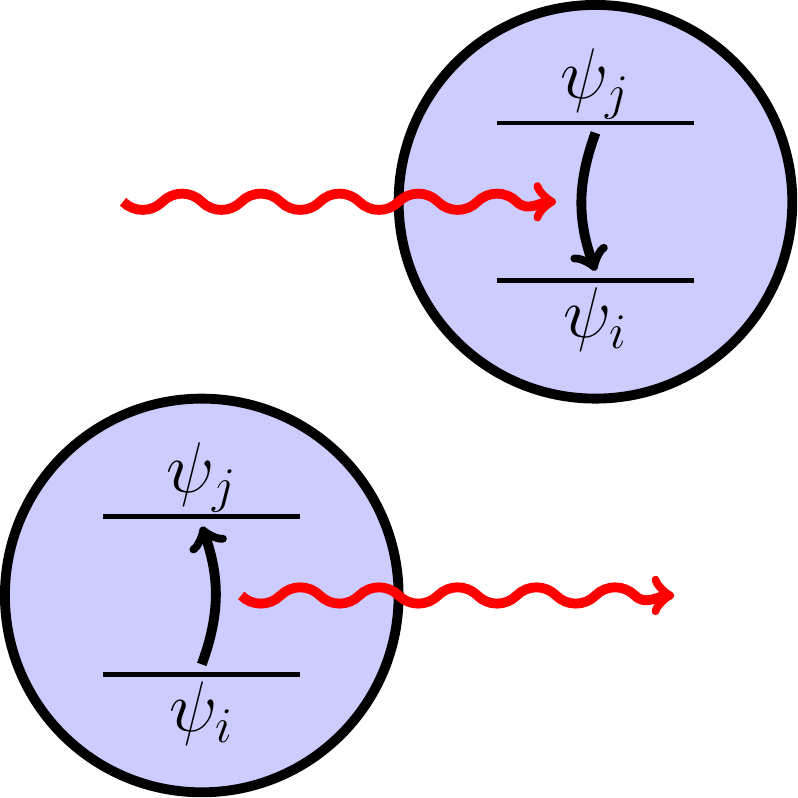}
}
\caption{
\label{fig:trans}
A microscopic picture of the momentum-independent couplings between the system and the environment in Eq.~(\ref{QBM}).
(a) The process of a particle decay from the high energy state $\lvert \psi_j \rangle$ to the low energy state $\lvert
\psi_i \rangle$ by emitting an energy quanta described by the term $V_{ijk} a_i^\dagger a_j b_k^\dagger$, and its
Hermitian conjugation process; (b) The process of the particle decay from the high energy state $\lvert \psi_j \rangle$
to the low energy state $\lvert \psi_i \rangle$ by absorbing an energy quanta described by the term $V_{ijk} a_i^\dagger
a_j b_k$, and its Hermitian conjugation process, which are physically infeasible.
}
\end{figure}

To have a clear physical picture on the microscopical level of interacting systems in the Feynman--Vernon model or the
Caldeira--Leggett model, it will be more convenient to reformulate the Hamiltonian of Eq.~(\ref{CLM}) in the second
quantization framework:
\begin{align}
    H_\text{CL}
    ={}
        & \frac{p^2}{2M}
        \! + \! V(x)
        \! + \! \sum_k \Bigl( \frac{p_k^2}{2m_k} + \frac{1}{2} m_k \omega_k^2 q_k^2 \Bigr) \notag \\
        & + \! \sum_k F_k(x) q_k
        \! + \! \sum_k F_k^2(x) / 2 m_k \omega_k^2 \notag \\
    ={}
        & \sum_i \varepsilon_i a_i^\dagger a_i
        \! + \!\! \sum_k \hbar \omega_k b_k^\dagger b_k
        \! + \!\! \sum_{ijk} V_{ijk} a_i^\dagger a_j (b_k^\dagger + b_k)
    .
\label{QBM}
\end{align}
Here, $\varepsilon_i$ is the spectrum of the system Hamiltonian $H_\textsc{s} = \frac{1}{2M} p^2 + V_\textsl{eff}(x)$
with $H_\textsc{s} \lvert \psi_i \rangle = \varepsilon_i \lvert \psi_i \rangle$, where $V_\textsl{eff}(x) = V(x) +
\sum_k \frac{F_k^2(x)}{2 m_k \omega_k^2}$. The operators $a_i^\dagger$ and $a_i$ are respectively the creation and
annihilation operators that create the system state $\lvert \psi_i \rangle$ from the vacuum state $\lvert 0 \rangle$:
$a_i^\dagger \lvert 0 \rangle = \lvert \psi_i \rangle$ and $a_i \lvert 0 \rangle = 0$. These creation and annihilation
operators obey the commutation or anti-commutation relation, $[a_i, a_j^\dagger]_\mp = a_i a_j^\dagger \mp a_j^\dagger
a_i = \delta_{ij}$, depending on the system being a bosonic or a fermionic system. The creation and annihilation
operators $b_k^\dagger$, $b_k$ of the harmonic bath can also be defined by $b_k^\dagger = \sqrt{\frac{m_k \omega_k}{2
\hbar}} (q_k - \frac{i}{m_k \omega_k} p_k)$, $b_k = \sqrt{\frac{m_k \omega_k}{2 \hbar}} (q_k + \frac{i}{m_k \omega_k}
p_k)$. The system-environment coupling $V_{ijk} = \sqrt{\frac{\hbar}{2 m_k \omega_k}} \langle \psi_i \rvert F_k(x)
\lvert \psi_j \rangle$ describes the transition of the system from the state $\lvert \psi_j \rangle$ to the state
$\lvert \psi_i \rangle$ due to the linear coupling to the environmental oscillating mode $\omega_k$. Now one can see a
physically troublesome process depicted by the linear system-environment coupling in the Feynman--Vernon model and the
Caldeira--Leggett model. That is, the system emits or absorbs a ``photon''
with the same energy results in the same transition of the system from the state $\lvert \psi_j \rangle$ to the state
$\lvert \psi_i \rangle$. This is quantum mechanically infeasible, as shown in Fig.~\ref{fig:trans}.
This seems to indicate that Eqs.~(\ref{FVM}) and (\ref{CLM}) are at least incomplete at quantum level if it is not
unphyscial.

In fact, a general Hamiltonian of the system coupled to the oscillator-bath environment should contain both the
position-dependent and the momentum-dependent system-environment couplings, as Caldeira and Leggett pointed out in their
original paper \cite{Leggett1983b},
\begin{align}
    H_\textsc{s-e}
    ={}
        & \frac{p^2}{2M}
        \! + \! V(x)
        \! + \! \sum_k \Bigl( \frac{p_k^2}{2m_k} + \frac{1}{2} m_k \omega_k^2 q_k^2 \Bigr) \notag \\
        & + \! \sum_k \bigl[ F_k(p,x) q_k + G_k(p,x) p_k \bigr]
        \! + \! \Phi(p,x)
    ,
\label{gQBM0}
\end{align}
where $\Phi(p,x)$ is a counter-term for the renormalization of system Hamiltonian arisen from the system-environment
couplings. Equation (\ref{gQBM0}) contains various position and momentum dependent system-environment couplings. The
second quantization form of the above Hamiltonian is given by
\begin{align}
    H_\textsc{s-e}
    ={}
        & \! \sum_i \varepsilon_i a_i^\dagger a_i
        \! + \! \sum_k \hbar \omega_k b_k^\dagger b_k \notag \\
        & + \! \sum_{ijk} \bigl( V_{ijk}' a_i^\dagger a_j b_k^\dagger + V_{ijk}'^* a_j^\dagger a_i b_k \bigr)
    .
\label{Hse}
\end{align}
Similarly, $\varepsilon_i$ is the spectrum of the system Hamiltonian $H_\textsc{s} = \frac{p^2}{2M} + V(x) + \Phi(p,x)$
with $H_\textsc{s} \lvert \psi_i \rangle = \varepsilon_i \lvert \psi_i \rangle$, the system-environment couplings
$V_{ijk}'$ are given by
\begin{subequations}
\begin{align}
    V_{ijk}'
    ={}
        & \sqrt{\frac{\hbar}{2m_k\omega_k}}
        \langle \psi_i \rvert F_k(p,x) \lvert \psi_j \rangle \notag \\
        & \qquad\qquad + i \sqrt{\frac{\hbar m_k \omega_k}{2}}
        \langle \psi_i \rvert G_k(p,x) \lvert \psi_j \rangle
    .
\end{align}
\end{subequations}
Now it shows that the system-environment coupling (the last two terms in Eq.~(\ref{Hse})) contains the physical process
of the system transiting from the state $\lvert \psi_j \rangle$ to the state $\lvert \psi_i \rangle$ by
\textit{emitting} a photon, and the inverse process of transiting the state $\lvert \psi_i \rangle$ back to the state
$\lvert \psi_j \rangle$ by \textit{absorbing} a photon with the same energy. In other words, the original ambiguity in
the Feynman-Vernon model and in the Caldeira-Leggett model can be diminished when a momentum-dependent system-reservoir
coupling co-exists.

The above analysis indicates that although Eqs.~(\ref{FVM}) and (\ref{CLM}) are well-defined in classical mechanics,
their quantum mechanical version is somehow inconsistent and contains some physically infeasible processes.
But the inconsistency can be eliminated when a momentum-dependent system-environment coupling is included, as it is
shown above. In fact, for atoms interacting with radiative electromagnetic field, the momentum-dependent coupling is
naturally manifested. Explicitly, the non-relativistic atom-photon interaction Hamiltonian can be rigorously written as
\cite{Tannoudji1998}
\begin{align}
    H_\textsl{e-p}
    =
        \! \sum_i \frac{1}{2m} \bigl[ \bm{P}_i \! - \! e \bm{A}(\bm{r}_i) \bigr]^2
        \!\! + \! \sum_{\mathclap{\bm{k}, \lambda=1,2}} \hbar \omega_{\bm{k}}
        b_{\bm{k},\lambda}^\dagger b_{\bm{k},\lambda}
        \! + \! V_\textsl{coul}
    ,
\label{Hqed}
\end{align}
where $m$ and $e$ are the electron mass and charge, $\bm{P}_i = - i\hbar \bm{\nabla}_i$ is the $i$-th electron momentum
operator in the atom. The vector field $\bm{A}(\bm{r})$ is the canonical coordinate of the electromagnetic field, and is
quantized by
\begin{align}
    \bm{A}(\bm{r})
    =
        \! \sum_{\mathclap{\bm{k}, \lambda=1,2}}
        \, \sqrt{\frac{\hbar}{2 \epsilon_0 \omega_{\bm{k}} V}}
        \hat{\bm{e}}_\lambda
        \Bigl(
            b_{\bm{k},\lambda} e^{i \bm{k} \cdot \bm{r}}
            + b_{\bm{k},\lambda}^\dagger e^{- i \bm{k} \cdot \bm{r}}
        \Bigr)
\end{align}
with the Coulomb gauge $\bm{k} \cdot \hat{\bm{e}}_\lambda = 0$. The last term $V_\textsl{coul} =
V_\textsl{coul}^\textsl{e-N} + V_\textsl{coul}^\textsl{e-e}$ in Eq.~(\ref{Hqed}) contains all the electron-nuclei and
electron-electron Coulomb interactions in the atom arisen from the charge-induced scalar potentials. The atom-light
interaction in Eq.~(\ref{Hqed}) is given by the term $- \frac{e}{m} \sum_i \bm{A}(\bm{r}_i) \cdot \bm{P}_i$, when the
Coulomb gauge $\bm{\nabla} \cdot \bm{A}(\bm{r}) = 0$ is used. This is a typical momentum-dependent coupling between the
system and the environment if the atom is considered as the system and the radiative field is treated as the
environment.

Taking the second quantization to the electron energy eigenfunctions $\psi_\alpha(\bm{r})$ of the atom under the
interaction with nuclei, we have $\big( \frac{1}{2m} \bm{P}^2 + V_\textsl{coul}^\textsl{e-N}(\bm{r}) \big)
\psi_\alpha(\bm{r}) = E_\alpha \psi_\alpha(\bm{r})$, where the nuclei is considered to be static. Then the atom-light
interaction Hamiltonian of Eq.~(\ref{Hqed}) can be rewritten as
\begin{align}
    H_\textsl{e-p}
    ={}
        & \! \sum_\alpha E_\alpha a_\alpha^\dagger a_\alpha
        \! + \! V_\textsl{coul}^\textsl{e-e}
        \! + \! \sum_{\mathclap{\bm{k}, \lambda=1,2}}
        \hbar \omega_{\bm{k}}
        b_{\bm{k},\lambda}^\dagger b_{\bm{k},\lambda} \notag \\
        & + \! \sum_{\mathclap{\alpha \beta \bm{k}, \lambda=1,2}}
        \bigl(
            V_{\alpha \beta \bm{k} \lambda} a_\alpha^\dagger a_\beta b_{\bm{k},\lambda}^\dagger
            + V_{\alpha \beta \bm{k} \lambda}^* a_\beta^\dagger a_\alpha b_{\bm{k},\lambda}
        \bigr)
    .
\label{Hqed1}
\end{align}
Here we ignore the term $\frac{e^2}{2m} \sum_i \bm{A}^2(\bm{r}_i)$ because it only gives shift to the radiative
frequencies for all the radiative field modes. The operators $a_\alpha^\dagger$ creates the single particle state
$\psi_\alpha(\bm{r})$ from the vacuum state. The transition amplitude $V_{\alpha \beta \bm{k} \lambda} = - \frac{e}{m}
\sqrt{\frac{\hbar}{2 \epsilon_0 \omega_{\bm{k}} V}} \int e^{- i \bm{k} \cdot \bm{r}} \psi_\alpha^*(\bm{r})
\hat{\bm{e}}_\lambda \cdot \bm{P} \psi_\beta(\bm{r}) \mathrm{d}^3 \bm{r}$ describes the photon-emitting transition from
the atomic state $\psi_\beta$ to the atomic state $\psi_\alpha$ under the constraints of energy, angular momentum and
parity conservations (the transition selection rules). The transition amplitude $V_{\alpha \beta \bm{k} \lambda}^*$
describes the photon-absorbing transition from the atomic state $\psi_\alpha$ to the atomic state $\psi_\beta$. In
quantum optics, one usually takes the dipole moment approximation ($e^{i \bm{k} \cdot \bm{r}} \simeq 1$) and treats the
atom as a two-level system, then the momentum-dependent atom-light coupling $- \frac{e}{m} \sum_i \bm{A}(\bm{r}_i) \cdot
\bm{P}_i$ is approximated to an electric dipole in an electric field $- \bm{d} \cdot \bm{E}$ which would induce the
counter-rotating terms as we discussed in the introduction. These counter-rotating terms arisen from the dipole
approximation is indeed artificial. A rigorous proof of no existence of counter-rotating terms can be found from the
fundamental quantum theory of light-matter interaction, i.e., the quantum electrodynamics within the framework of
quantum field theory \cite{Peskin1995}.

On the other hand, even for the Feynman--Vernon model Eq.~(\ref{FVM}) and the Caldeira--Leggett model of
Eq.~(\ref{CLM}), one cannot solve exactly the corresponding nonequilibrium decoherence dynamics in general. Only when
one takes the system also as a simple harmonic oscillator, $H_\textsc{s} = \frac{1}{2M} p^2 + \frac{1}{2} M \Omega^2
x^2$, and focuses on the strict linear coupling between the system and the environment, i.e. $U(x) = x$ in
Eq.~(\ref{FVM}) and $F_k(x) = C_k x$ in Eq.~(\ref{CLM}), can an exact master equation for the reduced density matrix of
the system be derived \cite{Haake1985,Hu1992,Karrlein1997}. In such circumstances, the simplified Feynman--Vernon model
or Caldeira--Leggett model still contains infeasible physical processes at quantum level arisen from the canonical
quantization. Explicitly, let us write the Hamiltonian of Eq.~(\ref{FVM}) or Eq.~(\ref{CLM}) for the system being a
harmonic oscillator linearly coupled to the oscillator bath in the particle number representation,
\begin{align}
    H
    \! ={}
        & \frac{p^2}{2M}
        \! + \! \frac{1}{2} M \Omega^2 x^2
        \! + \! \sum_k \Bigl(
            \frac{p_k^2}{2m_k} \! + \! \frac{1}{2} m_k \omega_k^2 q_k^2
        \Bigr) \notag \\
        & \qquad + \! \sum_k C_k x q_k
        \! + \! \sum_k C_k^2 x^2 / 2 m_k \omega_k^2 \notag \\
    ={}
        & \hbar \omega_\textsc{s} a^\dagger a
        \! + \! \sum_k \hbar \omega_k b_k^\dagger b_k \notag \\
        & \qquad \! + \! \sum_k \! \hbar V_k (
            a^\dagger b_k \! + \! b_k^\dagger a
            \! + \! a^\dagger b_k^\dagger \! + \! b_k a
        )
    ,
\label{CLH}
\end{align}
where $\omega_\textsc{s}^2 = \Omega^2 + \Omega_\textsl{c}^2$ and $\Omega_\textsl{c}^2 = \sum_k C_k^2 / M m_k \omega_k^2$
is arisen from the counter-term to cancel the possible divergence of the renormalized frequency-shift
\cite{Leggett1983b}, and $V_k = C_k / 2 \sqrt{M \omega_\textsc{s} m_k \omega_k} $. As we have pointed out in the
introduction, physically, the terms proportional to $a^\dagger b_k^\dagger + b_k a$ correspond to the processes of
generating and annihilating two energy quanta separately from nothing.
Consider a practical realization of quantum harmonic oscillator by photons, these processes correspond to two-photon
creation and annihilation in nonlinear optics. These two-photon processes could be physically realized, for example, by
spontaneous parametric down-conversion which is very different from the other two terms proportional to $a^\dagger b_k +
b_k^\dagger a$ in Eq.~(\ref{CLH}). The later is a linear particle process corresponding to the quantum energy tunneling
(energy quanta exchange) between the system and the environment. Consequently, the system-environment coupling strengths
for the linear particle exchange and the nonlinear particle pairing production/annihilation must be different in the
reality, and therefore Eq.~(\ref{CLH}) is physically incomplete at quantum level.

However, if the momentum-dependent couplings are included, the general strict linear coupling Hamiltonian can have the
form
\begin{align}
    & \sum_k \bigl(
        C_{1k} x q_k \! + \! C_{2k} p p_k \! + \! C_{3k} x p_k \! + \! C_{4k} p q_k
    \bigr) \notag \\
    & \qquad =
        \sum_k \hbar \bigl(
            V_k a^\dagger b_k \! + \! V_k^* b_k^\dagger a
            \! + \! W_k a^\dagger b_k^\dagger \! + \! W_k^* b_k a
        \bigr)
    ,
\end{align}
which can resolve the above problem. In other words, the quantum mechanically feasible QBM Hamiltonian that includes
momentum-dependent couplings should be
\begin{align}
    H_\textsl{tot}
    ={}
        & \hbar \omega_\textsc{s} a^\dagger a \! + \! \sum_k \hbar \omega_k b_k^\dagger b_k
        \! + \! \sum_k \hbar (V_k a^\dagger b_k \! + \! V_k^* b_k^\dagger a) \! \notag \\
        & \qquad\qquad\qquad + \! \sum_k \hbar (W_k a^\dagger b_k^\dagger \! + \! W_k^* b_k a)
    ,
\label{mCLH}
\end{align}
where $W_k \neq V_k$ in general. Obviously, the conventional QBM Eq.~(\ref{CLH}) in the literature is a special case of
Eq.~(\ref{mCLH}) with $W_k = V_k$. However, Eq.~(\ref{mCLH}) can be easily realized in many different quantum systems. A
typical example is the cavity coupled to the environment with an external squeezed driving,
\begin{align}
    H
    ={}
        & \hbar \omega_0 c^\dagger c + f(t) c^{\dagger 2} + f^*(t) c^2 \notag \\
        & + \! \sum_k \hbar \omega_k b_k^\dagger b_k
        \! + \! \sum_k (
            \mathcal{V}_k c^\dagger b_k
            \! + \! \mathcal{V}_k^* b_k^\dagger c
        )
    .
\end{align}
Taking a Bogoliubov transformation $a = u c + v c^\dagger$, $a^\dagger = u^* c^\dagger + v^* c$ with $\lvert u \rvert^2
- \lvert v \rvert^2 = 1$, it is easy to show that this Hamiltonian becomes Eq.~(\ref{mCLH}) with $\omega_\textsc{s}^2 =
\omega_0^2 - 4 \lvert f(t) \rvert^2 / \hbar^2$ and $V_k = u \mathcal{V}_k / \hbar$, $W_k = - v \mathcal{V}_k^* / \hbar$
such that $W_k \neq V_k$. This shows explicitly how the system-environment coupling strengths for one-photon processes
and two-photon processes become different in the reality. Thus, the generalization of the QBM with momentum-dependent
coupling between the system and the environment is physically more acceptable.

Very recently, investigations of quantum photonics has made a great progress for solving boson sampling problem
\cite{Aaronson2011,Crespi2013,Zhong2020,Zhong2021}. Quantum photonics is made of many squeezers and interferometers
(beam splitters and phase shifters), which can be described by the following Hamiltonian including various photon loss,
\begin{align}
    H_\text{Q.P.}(t)
    ={}
        & \! \sum_i \bigl[
            \hbar \omega_{\textsc{s}, i} c_i^\dagger c_i
            \! + \! f_i(t) c_i^{\dagger 2}
            \! + \! f_i^*(t) c_i^2
        \bigr] \notag \\
        & + \! \sum_{ij} \bigl(
            B_{ij} c_i^\dagger c_j + B_{ij}^* c_j^\dagger c_i
        \bigr)
        + \! \sum_{\alpha k} \hbar \omega_{\alpha k} b_{\alpha k}^\dagger b_{\alpha k} \notag \\
        & + \! \sum_{i \alpha k} \bigl(
                \mathcal{V}_{i,\alpha k} c_i^\dagger b_{\alpha k}
                \! + \! \mathcal{V}_{i,\alpha k}^* b_{\alpha k}^\dagger c_i
        \bigr)
    ,
\end{align}
where index $\alpha$ labels different photon loss environments. After diagonalizing the Hamiltonian of the squeezers and
interferometers (the first two terms in the above Hamiltonian) through a Bogoliubov transformation, the total
Hamiltonian can be reduced to
\begin{align}
    H_\text{Q.P.}(t)
    =
        & \sum_i \hbar \omega_{\textsc{s},i}(t) a_i^\dagger a_i
        \! + \! \sum_{\alpha k} \hbar \omega_{\alpha k} b_{\alpha k}^\dagger b_{\alpha k} \notag \\
        & + \! \sum_{i \alpha k} \Bigl(
            V_{i,\alpha k}(t) a_i^\dagger b_{\alpha k}
            \! + \! V_{i,\alpha k}^*(t) b_{\alpha k}^\dagger a_i \notag \\
            & \qquad\; + W_{i,\alpha k}(t) a_i^\dagger b_{\alpha k}^\dagger
            \! + \! W_{i,\alpha k}^*(t) b_{\alpha k} a_i
        \Bigr)
    .
\label{mmCLH}
\end{align}
This is an extension of the generalized QBM Eq.~(\ref{mCLH}) to multimodes that is practically used in the integrated
quantum photonics for the current investigation of photonics quantum computing
\cite{Politi2009,Paesani2019,Elshaari2020,Arrazola2021,Liu2022}.

Analogically, in the current investigation of topological phases of matter, many quantum devices are made by
superconductor-semiconductor, superconductor-insulator and superconductor-metal hybrid systems
\cite{Fu2008,Sau2010,Oreg2010,Lutchyn2018}. They are coupled to leads and thereby become open quantum systems. The
corresponding Hamiltonian, after a Bogoliubov transformation, can be equivalently written as
\begin{align}
    H_\text{T.S.}
    ={}
        & \! \sum_i \varepsilon_i(t) a_i^\dagger a_i
        + \sum_{\alpha k} \epsilon_{\alpha k}(t) b_{\alpha k}^\dagger b_{\alpha k} \notag \\
        & + \! \sum_{i \alpha k} \Bigl(
            V_{i,\alpha k}(t) a_i^\dagger b_{\alpha k}
            \! + \! \Delta_{i,\alpha k}(t) a_i^\dagger b_{\alpha k}^\dagger
            \! + \! \text{H.c.}
        \Bigr)
    ,
\label{Hsuper}
\end{align}
where index $\alpha$ labels different leads. This Hamiltonian has the same structure as the one for photonic quantum
processor of Eq.~(\ref{mmCLH}), except that here it describes electron dynamics rather than photon dynamics in open
quantum systems. Physically, Eq.~(\ref{Hsuper}) can be considered as a fermionic realization of the generalized QBM. The
general exact master equation of Eq.~(\ref{Hsuper}) has been derived recently by us for initially correlated
system-environment states \cite{Huang2020}. Applications to the study of decoherence dynamics on exotic states, such as
anyons and Majorana quasiparticles for topological quantum computer, have also been explored
\cite{Lai2018,Lai2020,Yao2020,Xiong2021}.

The above analysis indicates that there has been in the literature too much emphasis on quantization (i.e. general
methods of obtaining quantum mechanics from classical methods) as opposed to the converse problem of the classical limit
of quantum mechanics. This is unfortunate because the latter is an important question for various areas of modern
physics while the former is a chimera \cite{Simon1980}. From a more fundamental point of view, since the system plus the
environment is considered as a closed system, the quantum physical process should ensure the compatibility between the
unitarity and the conservation of energy-momentum transfer in every single microscopic quantum process.
In the following, we shall focus on the more general and quantum mechanically self-consistent model with
Eq.~(\ref{mCLH}) as a generalization of the conventional QBM. We will derive its exact master equation for both
initially decoupled and initially entangled system-environment states. For its applications in quantum photonics, such
as the integrated quantum processor made of many squeezers and interferometers with hundred more photonic modes for
photonics quantum computing that is described by Eq.~(\ref{mmCLH}), we will leave it for further investigation.

\section{The Exact Master Equation of the generalized QBM}
\label{sec:EME}
For an arbitrary initial state $\rho_\textsl{tot}(t_0)$, the time evolution of the total system (system plus
environment) is determined by the Liouville--von~Neumann equation of the quantum mechanics \cite{Neumann1955},
\begin{align}
    \frac{\mathrm{d}}{\mathrm{d} t} \rho_\textsl{tot}(t)
    =
        \frac{1}{i\hbar} [H_\textsl{tot}(t), \rho_\textsl{tot}(t)]
    .
\label{voneq}
\end{align}
Because the system and the environment together form a closed system, the Liouville--von~Neumann equation is the same as
the Schr\"{o}dinger equation of quantum mechanics for the evolution of pure quantum states. But the
Liouville--von~Neumann equation is more general because it is also valid for statistically mixed states, such as
Eqs.~(\ref{ipfs}) and (\ref{inist}), where the Schr\"{o}dinger equation is not applicable. The time evolution of the
system is determined by the reduced density matrix which is defined as the partial trace of the total density matrix
over all the environment states,
\begin{align}
    \rho_\textsc{s}(t) = \operatorname{Tr}_\textsc{e}[\rho_\textsl{tot}(t)]
    .
\label{rdm}
\end{align}

Applying the partial trace over the environment states to Eq.~(\ref{voneq}) with the total Hamiltonian of
Eq.~(\ref{mCLH}), we obtain formally the exact master equation
\begin{align}
    & \frac{\mathrm{d}}{\mathrm{d} t} \rho_\textsc{s}(t)
    =
        \frac{1}{i\hbar} [H_\textsc{s}, \rho_\textsc{s}(t)]
        + \mathcal{L}^+[\rho_\textsc{s}(t)]
        + \mathcal{L}^-[\rho_\textsc{s}(t)]
    ,
\end{align}
where
\begin{subequations}
\label{supercurrent}
\begin{align}
    \mathcal{L}^+[\rho_\textsc{s}(t)]
    ={}
        & a^\dagger A[\rho_\textsc{s}(t)] + A^\dagger[\rho_\textsc{s}(t)] a
    , \\
    \mathcal{L}^-[\rho_\textsc{s}(t)]
    ={}
        & - a A^\dagger[\rho_\textsc{s}(t)] - A[\rho_\textsc{s}(t)] a^\dagger
    ,
\end{align}
\end{subequations}
and $A[\rho_\textsc{s}(t)]$ is a collective operator defined by
\begin{align}
    A[\rho_\textsc{s}(t)]
    \equiv
        - i \operatorname{Tr}_\textsc{e} \bigl[ \! \bigl(
            \bm{V} \! \cdot \! \bm{b}
            \! + \! \bm{W} \! \cdot \! \bm{b}^\dagger
        \bigr) \rho_\textsl{tot}(t) \bigr]
    .
\label{part_tr}
\end{align}
Here $\mathcal{L}^\pm[\rho_\textsc{s}(t)]$ are positive/negative supercurrent operators that describe particles flowing
into/out of the system. It contains the usual current through particle tunnelings between the system and environments,
and also the current via particle pair production and annihilation. The latter is similar to the Andreev reflections
through the interface between superconductors and normal metals. Particle dissipation and fluctuations arisen from the
system-environment back-actions can be naturally manifested after completed the partial trace over all the environment
states in Eq.~(\ref{part_tr}).

Now the problem left is how to carry out explicitly the partial trace over all the environment states in
Eq.~(\ref{part_tr}). For the initial total density matrix $\rho_\textsl{tot}(t_0)$ being a partition-free thermal state
(containing the initial system-environment correlation),
\begin{align}
    \rho_\textsl{tot}(t_0)
    =
        \frac{1}{Z_\textsl{tot}} \exp \bigl( - H_\textsl{tot} / k_\textsl{B} T \bigr)
    ,
\label{ipfs}
\end{align}
because the total Hamiltonian is a bilinear function of particle creation and annihilation operators of the system and
environment, the total density matrix and the reduced density matrix always evolve in Gaussian states after a dynamical
quench. In this case, it is rather easy to trace over all the environment states with Gaussian integrals in the coherent
state representation. Such configuration has been presented in our previous derivation of the exact master equation for
dissipative topological systems \cite{Huang2020}. Below we will focus on cases with initial decoupled state. In the end
of the section, we will make a connection to the master equation with the above initial correlated state.

Thus, we now assume that the system and the environment are initially decoupled, and the environment is initially in a
thermal state \cite{Feynman1963,Leggett1983a},
\begin{align}
    \rho_\textsl{tot}(t_0)
    =
        \rho_\textsc{s}(t_0) \otimes \rho_\textsc{e}^{\textsl{th}}
    ,
    \quad \rho_\textsc{e}^{\textsl{th}}
    =
        \frac{1}{Z_\textsc{e}} e^{ \! - \! H_\textsc{e} / k_\textsl{B} T}
    .
\label{inist}
\end{align}
Such initial state can be prepared by tuning coupling strength between the system and the environment to zero at initial
time. Although the initial total state is a system-environment decoupled state, the system initial state
$\rho_\textsc{s}(t_0)$ can be arbitrary (non-Gaussian states). Therefore, the partial trace over the environment states
in Eq.~(\ref{part_tr}) is rather difficult to carry out. To complete this partial trace, we shall use the coherent state
path integral method \cite{Zhang1990}. In the coherent state representation, the matrix element of the collective
operator Eq.~(\ref{part_tr}) can be expressed as
\begin{align}
    \langle z_t^* \rvert
    \!\! \begin{bmatrix}
        A[\rho_\textsc{s}(t)] \\ A^\dagger[\rho_\textsc{s}(t)]
    \end{bmatrix}
    \!\! \lvert z_t' \rangle
    \! ={}
        & \!\!\! \int \! \mathrm{d} \mu(z_0^*, z_0) \mathrm{d} \mu(z_0'^* \! , z_0')
        \langle z_0^* \rvert \rho_\textsc{s}(t_0) \lvert z_0' \rangle \notag \\
        & \qquad \times \! \mathcal{J}^{\!A}(z_t^*, z_t', t; z_0, z_0'^* \! , t_0)
    .
\label{Apropf}
\end{align}
In Eq.~(\ref{Apropf}), $\lvert z \rangle \equiv e^{a^\dagger z} \lvert 0 \rangle$ is the unnormalized coherent state.
Correspondingly, the integral measure over the complex space is given by $\mathrm{d} \mu(z^*, z) \equiv \frac{\mathrm{d}
z^* \mathrm{d} z}{2 \pi i} e^{- \lvert z \rvert^2}$. The $A$-operator associated propagating function
$\mathcal{J}^{\!A}(z_t^*, z_t', t; z_0, z_0'^* \! , t_0)$ is defined in a similar way as the propagating function for
the reduced density matrix in the coherent state representation \cite{Jin2010},
\begin{align}
    \langle z_t^* \rvert \rho_\textsc{s}(t) \lvert z_t' \rangle
    ={}
        & \!\!\! \int \! \mathrm{d}\mu(z_0^*, z_0) \mathrm{d}\mu(z_0'^* \! , z_0')
        \langle z_0^* \rvert \rho_\textsc{s}(t_0) \lvert z_0' \rangle \notag \\
        & \qquad\qquad \times \! \mathcal{J}(z_t^*, z_t', t; z_0, z_0'^* \! , t_0)
    .
\label{Rpropf}
\end{align}
The propagating function $\mathcal{J}(z_t^*, z_t', t; z_0, z_0'^* \! , t_0)$ fully describes the time evolution of the
reduced density matrix of Eq.~(\ref{rdm}). Similarly, $\mathcal{J}^{\!A}(z_t^*, z_t', t; z_0, z_0'^* \! , t_0)$ fully
determine the evolution of the collective operator $A[\rho_\textsc{s}(t)]$ of Eq.~(\ref{part_tr}).

Utilizing the coherent state path integrals, we have
\begin{align}
    & \mathcal{J}(z_t^*, z_t', t; z_0, z_0'^* \! , t_0)
    =
        \!\! \int_{z_0}^{z_t^*}
        \!\!\! \mathfrak{D}[z^* \! , z] e^{\frac{i}{\hbar} S_\textsc{s}[z^* \! , z]} \notag \\
        & \qquad\; \times \!\! \int_{z_0'^*}^{z_t'}
        \!\! \mathfrak{D}[z'^* \! , z'] e^{- \frac{i}{\hbar} S_\textsc{s}^*[z'^* \! , z']}
        \mathcal{F}[z^*, z'^* \! , z, z']
    ,
\label{dpropa}
\end{align}
and similarly,
\begin{align}
    & \mathcal{J}^{\!A}(z_t^*, z_t', t; z_0, z_0'^* \! , t_0)
    =
        \!\! \int_{z_0}^{z_t^*}
        \!\!\! \mathfrak{D}[z^* \! , z] e^{\frac{i}{\hbar} S_\textsc{s}[z^* \! , z]} \notag \\
        & \qquad\; \times \!\! \int_{z_0'^*}^{z_t'}
        \!\! \mathfrak{D}[z'^* \! , z'] e^{- \frac{i}{\hbar} S_\textsc{s}^*[z'^* \! , z']}
        \mathcal{F}^A[z^*, z'^* \! , z, z']
    .
\label{Adpropa}
\end{align}
Here, $S_\textsc{s}[z^*, z]$ is the action of the system, given by
\begin{align}
    & \frac{i}{\hbar} S_\textsc{s}[z^*, z]
    =
        \frac{1}{2} (z^*(t_0) z_0 + z_t^* z(t)) \notag \\
        & \qquad + \!\! \int_{t_0}^t \!\! \mathrm{d} \tau \Bigl\{
            \frac{\dot{z}^*(\tau) z(\tau) \! - \! z^*(\tau) \dot{z}(\tau)}{2}
            \! - \! i \omega_\textsc{s} z^*(\tau) z(\tau)
        \Bigr\}
    .
\end{align}
The first two terms in the action are geometric terms arisen from the boundary (the end points of the coherent state
path integral). This is because the paths in the coherent state path integrals are only fixed on one side: $z(t_0) =
z_0$, $z^*(t) = z_t^*$ and $z'^*(t_0) = z_0'^*$, $z'(t) = z_t'$, see Eqs.~(\ref{Apropf}) and (\ref{Rpropf}). While
$(z(t), z^*(t))$ and $(z'(t), z'^*(t))$ are not the complex conjugate pairs of the above boundary conditions, and they
must be determined by solving the path integral explicitly \cite{Faddeev1980}.

On the other hand, $\mathcal{F}[z^*, z'^* \! , z, z']$ in Eq.~(\ref{dpropa}) is the influence functional arisen from the
partial trace over all the environmental states \cite{Feynman1963}. It is obtained by rigorously integrating out all the
environment degrees of freedom through the closed-loop path integrals \cite{Lei2012},
\begin{align}
    & \mathcal{F}[z^*, z'^* \! , z, z'] \notag \\
    & \quad =
        \exp \biggl\{
            \! - \frac{1}{2} \! \int_{t_0}^t \!\! \mathrm{d} \tau \biggl\{
                \! \int_{t_0}^\tau \!\!\! \mathrm{d} \tau'
                \! \begin{bmatrix} z^{\textsl{q}*}(\tau) & z^\textsl{q}(\tau) \end{bmatrix}
                \! \mathcal{G}(\tau, \tau')
                \! \begin{bmatrix} z(\tau') \\ z^*(\tau') \end{bmatrix} \notag \\
                & \quad\quad - \! \int_\tau^t \!\! \mathrm{d} \tau'
                \! \begin{bmatrix} z'^*(\tau) & z'(\tau) \end{bmatrix}
                \! \mathcal{G}(\tau, \tau')
                \! \begin{bmatrix} z^\textsl{q}(\tau') \\ z^{\textsl{q}*}(\tau') \end{bmatrix} \notag \\
                & \quad\quad + \! \int_{t_0}^t \!\! \mathrm{d} \tau'
                \! \begin{bmatrix} z^{\textsl{q}*}(\tau) & z^\textsl{q}(\tau) \end{bmatrix}
                \! \widetilde{\mathcal{G}}(\tau, \tau')
                \! \begin{bmatrix} z^\textsl{q}(\tau') \\ z^{\textsl{q}*}(\tau') \end{bmatrix}
            \! \biggr\}
        \! \biggr\}
    ,
\label{rdmIF}
\end{align}
where we have introduced the new variables $z^\textsl{q} \equiv z - z'$, $z^{\textsl{q}*} \equiv z^* - z'^*$. The
integral kernels $\mathcal{G}(\tau,\tau')$ and $\widetilde{\mathcal{G}}(\tau,\tau')$ are the two-time system-environment
correlation functions,
\begin{subequations}
\label{setcf}
\begin{align}
    \mathcal{G}(\tau,\tau')
    & =
        \!\! \sum_k
        \! \begin{bmatrix}
            V_k(\tau\!-\!t_0) & \!\! W_k(\tau\!-\!t_0) \\
            W_k^*(\tau\!-\!t_0) & \!\! V_k^*(\tau\!-\!t_0)
        \end{bmatrix} \notag \\
        & \qquad \times \!\! \begin{bmatrix} 1 & 0 \\ 0 & -1 \end{bmatrix} \!\! \begin{bmatrix}
            V_k^*(\tau'\!-\!t_0) & \!\! W_k(\tau'\!-\!t_0) \\
            W_k^*(\tau'\!-\!t_0) & \!\! V_k(\tau'\!-\!t_0)
        \end{bmatrix}
    ,
\label{setc} \\
    \widetilde{\mathcal{G}}(\tau,\tau')
    & =
        \!\! \sum_k
        \! \begin{bmatrix}
            V_k(\tau\!-\!t_0) & \!\! W_k(\tau\!-\!t_0) \\
            W_k^*(\tau\!-\!t_0) & \!\! V_k^*(\tau\!-\!t_0)
        \end{bmatrix}
        \!\! \begin{bmatrix}
            \overline{n}_k(t_0) & \overline{s}_k(t_0) \\
            \overline{s}_k^*(t_0) & \overline{h}_k(t_0)
        \end{bmatrix} \notag \\
        & \qquad \times \!
        \! \begin{bmatrix}
            V_k^*(\tau'\!-\!t_0) & \!\! W_k(\tau'\!-\!t_0) \\
            W_k^*(\tau'\!-\!t_0) & \!\! V_k(\tau'\!-\!t_0)
        \end{bmatrix}
    .
\label{idsetc}
\end{align}
\end{subequations}
Here we have also introduced the notations $V_k(\tau-t_0) \equiv V_k e^{- i \omega_k (\tau - t_0)}$ and $W_k(\tau-t_0)
\equiv W_k e^{+ i \omega_k (\tau - t_0)}$.
The phase factors come from the particle evolution in the environment before it transits into the system through the
tunneling or is created from the pairing coupling in Eq.~(\ref{mCLH}). It shows that only the time correlation function
$\widetilde{\mathcal{G}}(\tau,\tau')$ relates to the initial environment state through the initial particle occupations
and particle squeezed parameters in the environment, $\overline{n}_k(t_0) = \langle b_k^\dagger(t_0) b_k(t_0) \rangle$,
$\overline{s}_k(t_0) = \langle b_k(t_0) b_k(t_0) \rangle$, $\overline{h}_k(t_0) = \langle b_k(t_0) b_k^\dagger(t_0)
\rangle = 1 + \overline{n}_k(t_0)$. For the initial environment thermal state Eq.~(\ref{inist}), the particle squeezed
parameters $\overline{s}_k(t_0)$ become zero but here we formally keep it so that it can also apply to other initial
environment states. These time correlations depict back-action processes between the system and the environment.
While, the operator $A$-associated influence functional, after taking the partial trace over the environment states, can
be reduced as
\begin{align}
    \mathcal{F}^A[z^*, z'^* \! , z, z']
    & =
        \! - \!\! \int_{t_0}^t \!\!\! \mathrm{d} \tau \biggl\{
            \mathcal{Z} \mathcal{G}(t,\tau)
            \! \begin{bmatrix} z(\tau) \\ z^*(\tau) \end{bmatrix} \notag \\
            & + \mathcal{Z} \widetilde{\mathcal{G}}(t,\tau)
            \!\! \begin{bmatrix} z^\textsl{q}(\tau) \\ z^{\textsl{q}*}(\tau) \end{bmatrix}
        \! \biggr\}
        \mathcal{F}[z^*, z'^* \! , z, z']
    ,
\end{align}
where $\mathcal{Z} = [ \begin{smallmatrix} 1 & 0 \\ 0 & -1 \end{smallmatrix} ]$ is the z-component Pauli matrix.

Furthermore, the path integrals in the propagating function of Eq.~(\ref{dpropa}) can be exactly carries out because of
quadratic form of the effective action (including the influence functional). The result is
\begin{align}
    & \mathcal{J}(z_t^*, z_t', t; z_0, z_0'^* \! , t_0)
    =
        \frac{1}{Z(t,t_0)} \exp \Bigl\{ z_0'^* z_0 + z_t^* z_t' \Bigr\} \notag \\
        & \quad \times \exp \biggl\{ \frac{1}{2} \biggl\{
            \! \begin{bmatrix} z_t^* & \! z_t' \end{bmatrix}
            \! \mathcal{Z}
            \! \begin{bmatrix} \! z^\textsl{q}(t) \! \\ \! z^{\textsl{q}*}(t) \! \end{bmatrix}
            \! - \! \begin{bmatrix} z_0'^* & \! z_0 \end{bmatrix}
            \! \mathcal{Z}
            \! \begin{bmatrix} \!z^\textsl{q}(t_0) \! \\ \! z^{\textsl{q}*}(t_0) \! \end{bmatrix}
        \! \biggr\} \! \biggr\}
    .
\label{solu_prop}
\end{align}
Here $z^\textsl{q}(t)$, $z^{\textsl{q}*}(t)$ and $z^\textsl{q}(t_0)$, $z^{\textsl{q}*}(t_0)$ are determined by the
stationary path,
\begin{subequations}
\label{uvteq}
\begin{align}
    & \frac{\mathrm{d}}{\mathrm{d} \tau}
    \! \begin{bmatrix} \! z^\textsl{q}(\tau) \! \\ \! z^{\textsl{q}*}(\tau) \! \end{bmatrix}
    \! + \! i \omega_\textsc{s} \mathcal{Z}
    \! \begin{bmatrix} \! z^\textsl{q}(\tau) \! \\ \! z^{\textsl{q}*}(\tau) \! \end{bmatrix}
    \! - \!\! \int_\tau^t \!\!\!\! \mathrm{d} \tau' \mathcal{Z} \mathcal{G}(\tau,\tau')
    \! \begin{bmatrix} \! z^\textsl{q}(\tau') \! \\ \! z^{\textsl{q}*}(\tau') \! \end{bmatrix}
    \! = 0
    , \\
    & \frac{\mathrm{d}}{\mathrm{d} \tau}
    \! \begin{bmatrix} z(\tau) \\ z^*(\tau) \end{bmatrix}
    \! + \! i \omega_\textsc{s} \mathcal{Z}
    \! \begin{bmatrix} z(\tau) \\ z^*(\tau) \end{bmatrix}
    \! + \!\! \int_{t_0}^\tau \!\!\!\! \mathrm{d} \tau' \mathcal{Z} \mathcal{G}(\tau,\tau')
    \! \begin{bmatrix} z(\tau') \\ z^*(\tau') \end{bmatrix} \notag \\
    & \qquad\qquad\qquad\qquad =
        \! - \!\! \int_{t_0}^t \!\!\!\! \mathrm{d} \tau'
        \mathcal{Z} \widetilde{\mathcal{G}}(\tau,\tau')
        \! \begin{bmatrix} z^\textsl{q}(\tau') \\ z^{\textsl{q}*}(\tau') \end{bmatrix}
    ,
\end{align}
\end{subequations}
with $t_0 \leq \tau \leq t$. This is a generalization of our previous work \cite{Lei2012} to the open systems with
paring correlations. To solve the above equation, we introduce the transformation with different boundary conditions of
the initial and final ends in the coherent state path integrals,
\begin{subequations}
\label{uvdecomp}
\begin{align}
    & \begin{bmatrix} z^\textsl{q}(\tau) \\ z^{\textsl{q}*}(\tau) \end{bmatrix}
    \! =
        \mathcal{U}(\tau,t)
        \! \begin{bmatrix} z^\textsl{q}(t) \\ z^{\textsl{q}*}(t) \end{bmatrix}
    , \\
    & \begin{bmatrix} z(\tau) \\ z^*(\tau) \end{bmatrix}
    \! =
        \mathcal{U}(\tau,t_0)
        \! \begin{bmatrix} z(t_0) \\ z^*(t_0) \end{bmatrix}
        \! - \! \mathcal{V}(\tau,t) \mathcal{Z}
        \! \begin{bmatrix} z^\textsl{q}(t) \\ z^{\textsl{q}*}(t) \end{bmatrix}
    .
\end{align}
\end{subequations}
Then, Eq.~(\ref{uvteq}) is reduced to
\begin{subequations}
\label{pGreenfs}
\begin{align}
    & \frac{\mathrm{d}}{\mathrm{d} \tau} \mathcal{U}(\tau,t_0)
    \! + \! i \omega_\textsc{s} \mathcal{Z} \mathcal{U}(\tau,t_0)
    \! + \!\! \int_{t_0}^\tau \!\!\! \mathrm{d} \tau'
    \mathcal{Z} \mathcal{G}(\tau,\tau') \mathcal{U}(\tau',t_0)
    = 0,
\label{uteq} \\
    & \frac{\mathrm{d}}{\mathrm{d} \tau} \mathcal{V}(\tau,t)
    \! + \! i \omega_\textsc{s} \mathcal{Z} \mathcal{V}(\tau,t)
    \! + \!\! \int_{t_0}^\tau \!\! \mathrm{d} \tau'
    \mathcal{Z} \mathcal{G}(\tau,\tau') \mathcal{V}(\tau',t) \notag \\
    & \qquad\qquad\qquad\qquad\qquad\quad =
        \! \int_{t_0}^t \!\! \mathrm{d} \tau'
        \mathcal{Z} \widetilde{\mathcal{G}}(\tau,\tau')
        \mathcal{U}(\tau',t) \mathcal{Z}
    ,
\label{vteq}
\end{align}
\end{subequations}
subjected to the boundary conditions $\mathcal{U}(t_0,t_0) = I$ and $\mathcal{V}(t_0,t) = 0$. It is interesting to find
that $\mathcal{U}(\tau,t_0)$ and $\mathcal{V}(\tau,t)$ are the generalization of the nonequilibrium Green functions for
open quantum systems we previously introduced \cite{Tu2008,Jin2010,Lei2012,Zhang2012} to the case involving pairing
processes. In terms of the usual definition of nonequilibrium Green functions, they can be expressed as
\begin{subequations}
\begin{align}
    & \mathcal{U}(t,t_0) \mathcal{Z}
    =
        \begin{bmatrix}
            \langle [a(t), a^\dagger(t_0)] \rangle &
            \langle [a(t), a(t_0)] \rangle \\
            \langle [a^\dagger(t), a^\dagger(t_0)] \rangle &
            \langle [a^\dagger(t), a(t_0)] \rangle
        \end{bmatrix}
    , \\
    & \mathcal{V}(\tau,t)
    =
        \begin{bmatrix}
            \langle a^\dagger(t) a(\tau) \rangle &
            \langle a(t) a(\tau) \rangle \\
            \langle a^\dagger(t) a^\dagger(\tau) \rangle &
            \langle a(t) a^\dagger(\tau) \rangle
        \end{bmatrix} \notag \\
        & \quad - \mathcal{U}(\tau,t_0)
        \! \begin{bmatrix}
            \langle a^\dagger(t_0) a(t_0) \rangle &
            \langle a(t_0) a(t_0) \rangle \\
            \langle a^\dagger(t_0) a^\dagger(t_0) \rangle &
            \langle a(t_0) a^\dagger(t_0) \rangle
        \end{bmatrix}
        \! \mathcal{U}^\dagger(t,t_0)
    .
\end{align}
\end{subequations}
They fully describe the transient dissipation and fluctuation dynamics of open quantum system incorporating various
back-actions from the environment, and obey the integro-differential equations of motion (\ref{pGreenfs}) which can also
be derived from the Heisenberg equation of motion, see the detailed derivation given in Appendix~\ref{sec:UV}. The
solution of Eq.~(\ref{vteq}) for the nonequilibrium noise-induced correlation Green function $\mathcal{V}(\tau,t)$ can
be obtained easily,
\begin{align}
    \mathcal{V}(\tau,t)
    =
        \!\! \int_{t_0}^\tau \!\!\! \mathrm{d} \tau'
        \!\! \int_{t_0}^t \!\!\! \mathrm{d} t'
        \mathcal{U}(\tau,\tau') \mathcal{Z}
        \widetilde{\mathcal{G}}(\tau',t') \mathcal{Z}
        \mathcal{U}^\dagger(t,t')
    .
\label{vtsol}
\end{align}
which is indeed the generalization of the Keldysh's correlation Green function with pairing couplings
\cite{Keldysh1965}. This solution is the generalized nonequilibrium fluctuation-dissipation theorem \cite{Zhang2012},
also see the derivation given in Appendix~\ref{sec:UV}.

With equation of motion Eq.~(\ref{uvteq}), the $A$-associated propagating functions become
\begin{align}
    \mathcal{J}^A(z_t^*, z_t', t; z_0, z_0'^* \! , t_0)
    ={}
        & \biggl\{ \!
            \begin{bmatrix} \dot{z}(t) \\ \dot{z}^*(t) \end{bmatrix}
            \! + \! i \omega_\textsc{s} \mathcal{Z}
            \begin{bmatrix} z(t) \\ z^*(t) \end{bmatrix}
        \! \biggr\} \notag \\
        & \times \! \mathcal{J}(z_t^*, z_t', t; z_0, z_0'^* \! , t_0)
    ,
\end{align}
and using Eq.~(\ref{uvdecomp}), the additional term can be fully determined in terms of the fixed end points of the path
integrals. From the above result, we obtain the current superoperators Eq.~(\ref{supercurrent}) in the coherent state
representation
\begin{subequations}
\label{supcu}
\begin{align}
    & \langle z^* \rvert \mathcal{L}^+[\rho_\textsc{s}(t)] \lvert z' \rangle
    \! = -
        \biggl\{
            \begin{bmatrix} z^* & z' \end{bmatrix}
            \! \mathcal{K}(t)
            \! \begin{bmatrix} \partial_{z^*} \!\\ z^* \! \end{bmatrix} \notag \\
            & \qquad\qquad\quad + \! \begin{bmatrix} z^* & z' \end{bmatrix}
            \! \Lambda(t)
            \! \begin{bmatrix} \partial_{z^*} \! - \! z' \\ \partial_{z'} \! - \! z^* \end{bmatrix}
        \biggr\}
        \langle z^* \rvert \rho_\textsc{s}(t) \lvert z' \rangle
    , \\
    & \langle z^* \rvert \mathcal{L}^-[\rho_\textsc{s}(t)] \lvert z' \rangle
    \! = \!
        \biggl\{
            \begin{bmatrix} \partial_{z'} & \partial_{z^*} \end{bmatrix}
            \! \mathcal{K}(t)
            \! \begin{bmatrix} \partial_{z^*} \\ z^* \end{bmatrix} \notag \\
            & \qquad\qquad\quad + \! \begin{bmatrix} \partial_{z'} & \partial_{z^*} \end{bmatrix}
            \! \Lambda(t)
            \! \begin{bmatrix} \partial_{z^*} \! - \! z' \\ \partial_{z'} \! - \! z^* \end{bmatrix}
        \biggr\}
        \langle z^* \rvert \rho_\textsc{s}(t) \lvert z' \rangle
    ,
\end{align}
\end{subequations}
where $\mathcal{K}(t)$ and $\Lambda(t)$ are defined as
\begin{subequations}
\label{renormph}
\begin{align}
    \mathcal{K}(t)
    \equiv{} &
        \! - i \omega_\textsc{s} \mathcal{Z}
        - \dot{\mathcal{U}}(t,t_0) \mathcal{U}(t,t_0)^{-1}
    , \\
    \Lambda(t)
    \equiv{} &
        \dot{\mathcal{V}}(t,t)
        - \dot{\mathcal{U}}(t,t_0) \mathcal{U}(t,t_0)^{-1} \mathcal{V}(t,t)
    .
\label{lamd}
\end{align}
\end{subequations}
Note that here $\dot{\mathcal{V}}(t,t)$ represents only the time-derivative with respect to the first parameter.

After some rearrangement, one can rewrite the exact master equation as the standard form as we did before
\begin{align}
    \frac{\mathrm{d}}{\mathrm{d} t} \rho_\textsc{s}(t)
    ={}
        & \frac{1}{i\hbar} \bigl[ H_\textsc{s}'(t), \rho_\textsc{s}(t) \bigr] \notag \\
        & + \gamma(t) \bigl\{
            a \rho_\textsc{s}(t) a^\dagger
            \! - \! \tfrac{1}{2} a^\dagger a \rho_\textsc{s}(t)
            \! - \! \tfrac{1}{2} \rho_\textsc{s}(t) a^\dagger a
        \bigr\} \notag \\
        & + \widetilde{\gamma}(t) \bigl\{
            a^\dagger \rho_\textsc{s}(t) a
            \! - \! \tfrac{1}{2} a a^\dagger \rho_\textsc{s}(t)
            \! - \! \tfrac{1}{2} \rho_\textsc{s}(t) a a^\dagger \notag \\
            & \qquad\quad + a \rho_\textsc{s}(t) a^\dagger
            \! - \! \tfrac{1}{2} a^\dagger a \rho_\textsc{s}(t)
            \! - \! \tfrac{1}{2} \rho_\textsc{s}(t) a^\dagger a
        \bigr\} \notag \\
        & - \overline{\gamma}(t) \bigl\{
            a^\dagger \rho_\textsc{s}(t) a^\dagger
            \! - \! \tfrac{1}{2} a^{\dagger 2} \rho_\textsc{s}(t)
            \! - \! \tfrac{1}{2} \rho_\textsc{s}(t) a^{\dagger 2}
        \bigr\} \notag \\
        & - \overline{\gamma}^*(t) \bigl\{
            a \rho_\textsc{s}(t) a
            \! - \! \tfrac{1}{2} a^2 \rho_\textsc{s}(t)
            \! - \! \tfrac{1}{2} \rho_\textsc{s}(t) a^2
        \bigr\}
    ,
\label{geme}
\end{align}
where
$
    H_\textsc{s}'(t)
    =
        \hbar \omega_{\textsc{s}}'(t) a^\dagger a
        + \frac{1}{2} \hbar \overline{\omega}_{\textsc{s}}'(t) a^{\dagger 2}
        + \frac{1}{2} \hbar \overline{\omega}_{\textsc{s}}'^*(t) a^2
$.
Other coefficients are given by
\begin{subequations}
\label{edcingeme}
\begin{align}
    & \omega_\textsc{s}'(t)
    =
        \omega_\textsc{s}
        - \frac{i}{2} [\mathcal{K}_{11}(t) - \mathcal{K}_{11}^*(t)]
    , \\
    & \overline{\omega}_\textsc{s}'(t)
    =
        - \frac{i}{2} [\mathcal{K}_{12}(t) + \mathcal{K}_{21}^*(t)]
    , \\
    & \gamma(t)
    =
        \mathcal{K}_{11}(t) + \mathcal{K}_{11}^*(t)
    , \\
    & \widetilde{\gamma}(t)
    =
        \Lambda_{11}(t) + \Lambda_{11}^*(t)
    , \\
    & \overline{\gamma}(t)
    =
        \Lambda_{12}(t) + \Lambda_{21}^*(t)
    .
\end{align}
\end{subequations}
It shows that the coefficients $\omega_\textsc{s}'(t)$ and $\overline{\omega}_\textsc{s}'(t)$ are the renormalized
frequency and renormalized pairing strength modified by the environment. By diagonalizing the renormalized Hamiltonian
with a Bogoliubov transformation, the renormalized eigen-frequency is $\omega_\textsl{r}(t) = \sqrt{\lvert
\omega_\textsc{s}'(t) \rvert^2 - \lvert \overline{\omega}_\textsc{s}'(t) \rvert^2}$. The term proportional to
$\gamma(t)$ represents a non-unitary evolution that describes the dissipation dynamics induced by the environment.
Notice that the definition of $\gamma(t)$ here is twice of that in our previous paper \cite{Huang2020} because we
rewritten the superoperator multiplied by a factor $1/2$ so that all the superoperators have formally the same standard
Lindblad form \cite{Lindblade}.
As a result, $\gamma(t)$ corresponds precisely to the decay coefficient of the system. Also, there is a term
proportional to $\gamma'(t)$ in our previous paper \cite{Huang2020} which vanishes here for the single Brownian particle
system. The other two terms proportional to $\widetilde{\gamma}(t)$ and $\overline{\gamma}(t)$ in the master equation
describe fluctuation (diffusion) dynamics. They depend on the initial state of the environment.
Note that the $\overline{\gamma}(t)$-associated fluctuations are induced by pair production and annihilation
between the system and the environment.

To understand further the physical picture of the renormalized energy, the dissipation and fluctuation dynamics induced
by the environment, let us calculate the mean values and their deviations (equivalent to the quadrature
covariance in terms of the position and momentum) of the system variables from the exact master equation.
From the exact master equation (\ref{geme}), it is easy to find that
\begin{align}
    \frac{\mathrm{d}}{\mathrm{d} t}
    \! \begin{bmatrix}
        \langle a(t) \rangle \\
        \langle a^\dagger(t) \rangle
    \end{bmatrix}
    \! =
        \! \begin{bmatrix}
            - i \omega_\textsc{s}'(t) \! - \! \gamma(t)/2 &
            \! - i \overline{\omega}_\textsc{s}'(t) \\
            i \overline{\omega}_\textsc{s}'^*(t) &
            \! i \omega_\textsc{s}'(t) \! - \! \gamma(t)/2
        \end{bmatrix}
        \!\! \begin{bmatrix}
            \langle a(t) \rangle \\
            \langle a^\dagger(t) \rangle
        \end{bmatrix}
    ,
\label{fvariables}
\end{align}
where $\langle a(t) \rangle = \operatorname{Tr}_\textsc{s}[a \rho_\textsc{s}(t)]$ and $\langle a^\dagger(t) \rangle =
\operatorname{Tr}_\textsc{s}[a^\dagger \rho_\textsc{s}(t)]$. The above equation shows that the Brownian particle
undergoes a damping oscillation with the renormalized frequencies $\omega_\textsc{s}'(t)$ and
$\overline{\omega}_\textsc{s}'(t)$ and the decay coefficient $\gamma(t)$ induced from the environment. The mean value
$\langle a(t) \rangle$ is mixed with its complex conjugation which is related to the squeezed term $\frac{1}{2} \hbar
\overline{\omega}_\textsc{s}'(t) a^{\dagger 2}$ in the renormalized Hamiltonian, due to the pairing coupling between the
system and the environment.

On the other hand, the mean-value deviations obey the following equations
\begin{align}
    & \frac{\mathrm{d}}{\mathrm{d} t}
    \! \begin{bmatrix}
        \Delta n(t) &
        \Delta s(t) \\
        \Delta s^*(t) &
        \Delta h(t)
    \end{bmatrix} \notag \\
    & \quad =
        \biggl\{
            \! \begin{bmatrix}
                - i \omega_\textsc{s}'(t) \! - \! \gamma(t)/2 &
                - i \overline{\omega}_\textsc{s}'(t) \\
                i \overline{\omega}_\textsc{s}'^*(t) &
                i \omega_\textsc{s}'(t) \! - \! \gamma(t)/2
            \end{bmatrix}
            \!\! \begin{bmatrix}
                \Delta n(t) &
                \Delta s(t) \\
                \Delta s^*(t) &
                \Delta h(t)
            \end{bmatrix} \notag \\
            & \quad\qquad + \text{H.c.}
        \biggr\}
        + \begin{bmatrix}
            \widetilde{\gamma}(t) &
            \overline{\gamma}(t) \\
            \overline{\gamma}^*(t) &
            \gamma(t) + \widetilde{\gamma}(t)
        \end{bmatrix}
    ,
\label{gcont}
\end{align}
where $\Delta n(t) = \operatorname{Tr}_\textsc{s}[\delta a^\dagger \delta a \rho_\textsc{s}(t)]$, $\Delta s(t) =
\operatorname{Tr}_\textsc{s}[\delta a \delta a \rho_\textsc{s}(t)]$, $\Delta h(t) = \operatorname{Tr}_\textsc{s}[\delta
a \delta a^\dagger \rho_\textsc{s}(t)] = 1 + \Delta n(t)$, and $\delta a = a - \langle a \rangle$. The first term and
its Hermitian conjugate in the right hand side of the above equation describe a damping oscillation of the quadrature
variables, which follows the same dynamics as given by Eq.~(\ref{fvariables}). The last term in Eq.~(\ref{gcont}) is the
noise sources coming from the particle transitions and pair productions between the system associated with the initial
state of the environment. Thus, $\widetilde{\gamma}(t)$ and $\overline{\gamma}(t)$ can be interpreted as the diffusion
rates.

We also find that the exact master equation (\ref{geme}) derived with the initial decoupled state has indeed the same
form as the exact master equation for the initial correlated density matrix $\rho_\textsl{tot}(t_0)$ of Eq.~(\ref{ipfs})
that we have derived in the previous work (i.e., Eq.~(12) in Ref.~\cite{Huang2020}). In Eq.~(\ref{ipfs}),
$H_\textsl{tot}$ is the total Hamiltonian of the system and the environment plus their interaction, as given by
Eq.~(\ref{mCLH}). Thus, the initial state cannot be decoupled into a direct product of the system state with the
environment state. Of course, to let the system and the environment evolve into a nonequilibrium evolution with such an
initial thermal state, one needs to quench the system. A simple quench can be made by replacing the system frequency
$\omega_\textsc{s}$ with time-dependent one $\omega_\textsc{s}(t)$, which is easy to be realized in experiments. All the
formulas derived in this section remain the same forms with such a quenching modulation. As one can find from
\cite{Huang2020}, with the initial correlated state Eq.~(\ref{ipfs}), the only modification in the exact master equation
is the fluctuation coefficients $\widetilde{\gamma}(t)$ and $\overline{\gamma}(t)$, which are still determined by the
nonequilibrium Green function $\mathcal{V}(t,t)$ but $\mathcal{V}(t,t)$ of Eq.~(\ref{renormph}) is modified by
\begin{align}
    \mathcal{V}(t,t)
    & =
        \!\! \int_{t_0}^t \!\!\! \mathrm{d} \tau
        \!\! \int_{t_0}^t \!\!\! \mathrm{d} \tau'
        \mathcal{U}(t,\tau) \mathcal{Z}
        \bigl[
            \widetilde{\mathcal{G}}(\tau,\tau')
            + \widetilde{\mathcal{G}}^{(\textsc{es})}(\tau,\tau') \notag \\
            & \qquad\qquad\qquad + \widetilde{\mathcal{G}}^{(\textsc{es})\dagger}(\tau',\tau)
        \bigr] \mathcal{Z}
        \mathcal{U}^\dagger(t,\tau')
    .
\end{align}
Here $\widetilde{\mathcal{G}}^{(\textsc{es})}(\tau,\tau')$ is given by
\begin{align}
    \widetilde{\mathcal{G}}^{(\textsc{es})}(\tau,\tau')
    ={}
        & \! - 2 i \! \sum_k
        \! \begin{bmatrix}
            V_k(\tau\!-\!t_0) & \!\! W_k(\tau\!-\!t_0) \\
            W_k^*(\tau\!-\!t_0) & \!\! V_k^*(\tau\!-\!t_0)
        \end{bmatrix} \notag \\
        & \! \times
        \! \begin{bmatrix}
            \overline{n}_k'(t_0) & \overline{s}_k'(t_0) \\
            \overline{s}_k'^*(t_0) & \overline{n}_k'^*(t_0)
        \end{bmatrix} \mathcal{Z}
        \delta(\tau' - t_0)
    ,
\end{align}
with $\overline{n}_k'(t_0) = \langle a^\dagger(t_0) b_k(t_0) \rangle$ and $\overline{s}_k'(t_0) = \langle a(t_0)
b_k(t_0) \rangle$ being initial correlations between the system and the environment. The factor $2$ comes from the
integral $\int_{t_0}^t \mathrm{d}\tau \delta(\tau - t_0) = 1/2$.


For the special case of $W_k = 0$,
Eq.~(\ref{mCLH}) is reduced to the Fano Hamiltonian that describes a discrete state coupling to a continuous spectral
bath. In this cases, the pairing-production or pairing-annihilation related coefficients vanish:
$\overline{\gamma}(t) = 0$ and $\overline{\omega}_\textsc{s}'(t) = 0$. Correspondingly, the exact master equation
(\ref{geme}) is reduced to the one given in our earlier works
\cite{Wu2010,Xiong2010,Lei2011,Lei2012,Xiong2015,Tang2011,Huang2020}. For another special case of $W_k = V_k$ which
corresponds to conventional QBM model \cite{Leggett1983a,Hu1992}, we will discuss the corresponding reduction in details
in the next section.

\section{Reproduction of the conventional QBM and the resolution of the initial jolt problem}
\label{sec:QBM}
In this section, we will show that the Hu-Paz-Zhang master equation of the QBM model \cite{Hu1992} is a special case of
our master equation Eq.~(\ref{geme}) for the generalized QBM. We will also show that the long-standing initial jolt
problem in the conventional QBM is artificial but it has nothing to do with the initial decoupled state.

Without including momentum-dependent couplings, the strength of the coupling and the pairing are the same $W_k = V_k$,
as we shown in Eq.~(\ref{CLH}). This condition leads to the following relations between the renormalized frequency,
dissipation (decay) and fluctuation (diffusion) coefficients
\begin{subequations}
\label{evwc}
\begin{align}
    & \operatorname{Re}[\overline{\omega}_\textsc{s}'(t)] = \omega_\textsc{s}'(t) - \omega_\textsc{s}
    , \\
    & \operatorname{Im}[\overline{\omega}_\textsc{s}'(t)] = \gamma(t)/2
    , \\
    & \operatorname{Re}[\overline{\gamma}(t)] = - \gamma(t)/2 - \widetilde{\gamma}(t)
    .
\end{align}
\end{subequations}
Through these relations, the Hu-Paz-Zhang master equation of the conventional QBM can be reproduced easily from our
general exact master equation Eq.~(\ref{geme}) as a special limit. It can be expressed alternatively as
\begin{align}
    i \hbar \frac{\mathrm{d}}{\mathrm{d} t} \rho_\textsc{s}(t)
    ={}
        & \! \bigl[ H_\textsl{R}(t), \rho_\textsc{s}(t) \bigr] \notag \\
        & + \Gamma(t) \bigl\{
            x \rho_\textsc{s}(t) p - \tfrac{1}{2} px \rho_\textsc{s}(t) - \tfrac{1}{2} \rho_\textsc{s}(t) px \notag \\
            & \qquad\quad
            - p \rho_\textsc{s}(t) x + \tfrac{1}{2} xp \rho_\textsc{s}(t) + \tfrac{1}{2} \rho_\textsc{s}(t) xp
        \bigr\} \notag \\
        & + i M \Gamma(t) h(t) \bigl\{
            2x \rho_\textsc{s}(t) x - x^2 \rho_\textsc{s}(t) - \rho_\textsc{s}(t) x^2
        \bigr\} \notag \\
        & - i \Gamma(t) f(t) \bigl\{
            x \rho_\textsc{s}(t) p - \tfrac{1}{2} px \rho_\textsc{s}(t) - \tfrac{1}{2} \rho_\textsc{s}(t) px \notag \\
            & \qquad\qquad
            + p \rho_\textsc{s}(t) x - \tfrac{1}{2} xp \rho_\textsc{s}(t) - \tfrac{1}{2} \rho_\textsc{s}(t) xp
        \bigr\}
    ,
\label{hpz}
\end{align}
where $H_\textsl{R}(t)$ is the complete renormalized system Hamiltonian,
\begin{align}
    H_\textsl{R}(t)
    \equiv{}
        & \tfrac{1}{2 M} p^2 + \tfrac{1}{2} M \omega_\textsc{s}^2 x^2 \notag \\
        & + \tfrac{1}{2} M \delta \omega_\textsc{s}^2(t) x^2
        + \tfrac{1}{2} \Gamma(t) (xp \! + \! px)
    .
\label{rsH}
\end{align}
All the coefficients in the above master equation are deduced from the coefficients of the general master equation
(\ref{geme}) under the condition $W_k = V_k$. The results are given by
\begin{subequations}
\begin{align}
    & \delta \omega_\textsc{s}^2(t)
    =
        2 \omega_\textsc{s} \operatorname{Re}[\overline{\omega}_\textsc{s}'(t)]
    ,
    \quad \Gamma(t)
    =
        \gamma(t)/2
    , \\
    & \Gamma(t) h(t)
    =
        \omega_\textsc{s} \operatorname{Re}[\overline{\gamma}(t)]
    ,
    \quad \Gamma(t) f(t)
    =
        - \operatorname{Im}[\overline{\gamma}(t)]
    .
\end{align}
\end{subequations}

Notice that in the original Hu-Paz-Zhang master equation, the renormalized system Hamiltonian is incomplete, because not
all commutator terms (associated with the unitary evolution) have been separated from the anti-commutator terms
(non-unitary evolution) in the master equation. This is how our complete renormalized system Hamiltonian is defined, as
given by Eq.~(\ref{rsH}). It shows that the renormalized system Hamiltonian contains a potential of the
position-momentum coupling. The reason that this momentum-dependent potential is misplaced into the dissipation (decay)
term in the Hu-Paz-Zhang equation is because at the special case $W_k = V_k$, we have
$\operatorname{Im}[\overline{\omega}_\textsc{s}'(t)] = \gamma(t)/2$, see Eq.~(\ref{evwc}), so that this
renormalization-induced potential has the same coefficient as the dissipation term in the master equation. On the other
hand, this renormalization-induced momentum-dependent potential also indicates that including the momentum-dependent
couplings between the system and the environment is a natural consequence, because even if only the position-position
coupling is considered as in the conventional QBM, the renormalization not only changes the frequency of the system, but
also induces a position-momentum coupling potential in the system Hamiltonian. As a consequence of this
position-momentum coupling potential, the physical frequency of the Brownian particle is $\omega_\textsl{p}^2(t) =
\omega_\textsc{s}^2 + \delta \omega_\textsc{s}^2(t) - \Gamma^2(t)$, which is consistent with the renormalized
eigen-frequency $\omega_\textsl{r}(t) = \sqrt{\lvert \omega_\textsc{s}'(t) \rvert^2 - \lvert
\overline{\omega}_\textsc{s}'(t) \rvert^2}$ in our general formalism, see the discussion after Eq.~(\ref{edcingeme}).

The decay and diffusion of the QBM can be manifested clearlier from the equation of motion for the quadrature covariance
variables, which can be found from the master equation (\ref{hpz}) directly:
\begin{align}
    & \frac{\mathrm{d}}{\mathrm{d} t}
    \begin{bmatrix}
        \langle \Delta x^2(t) \rangle &
        \langle\Delta xp(t) \rangle \\
        \langle\Delta px(t) \rangle &
        \langle \Delta p^2(t) \rangle
    \end{bmatrix} \notag \\
    & =
        \biggl\{
            \! \begin{bmatrix}
                0 &
                1/M \\
                - M (\omega_\textsc{s}^2 \!\! + \delta \omega_\textsc{s}^2(t)) &
                - 2 \Gamma(t)
            \end{bmatrix}
            \!\! \begin{bmatrix}
                \langle \Delta x^2(t) \rangle &
                \langle\Delta xp(t) \rangle \\
                \langle\Delta px(t) \rangle &
                \langle \Delta p^2(t) \rangle
            \end{bmatrix} \notag \\
            & \quad\quad + \text{H.c}
        \biggr\}
        + \begin{bmatrix}
            0 &
            \Gamma(t) f(t) \\
            \Gamma(t) f(t) &
            2 M \Gamma(t) h(t)
        \end{bmatrix}
    ,
\end{align}
where $\langle \Delta xp(t) \rangle = \frac{1}{2} \langle \{\Delta x(t), \Delta p(t)\} \rangle$. Thus, the coefficients
$\Gamma(t) f(t)$ and $\Gamma(t) h(t)$ can be interpreted as the diffusion rates of the quadrature covariance variables
$\langle \Delta xp(t) \rangle$ and $\langle \Delta p^2(t) \rangle$, respectively. The missing diffusion rate for
$\langle \Delta x^2(t) \rangle$ is due to the lack of position-momentum coupling in the conventional QBM Hamiltonian.

Based on the above reproduction of Hu-Paz-Zhang master equation from our generalized QBM, we would also like to address
a long-standing debate issue in the conventional QBM, called ``initial jolt'' \cite{Zurek1989,Hu1992,Ford2001}. The
initial jolt is associated with a peak at the beginning of the diffusion coefficient $\Gamma(t) h(t)$ at low temperature
limit in the Hu-Paz-Zhang master equation, which is sensitive to the cut-off frequency $\Lambda$ for the Ohmic-type
spectral density. The inverse of this cut-off frequency, $1/\Lambda$, is the typical time scale of the environment. The
amplitude of the initial jolt is proportional to $\Lambda$, and it diverges when one takes a constant damping
coefficient for the conventional QBM model. This corresponds to an Ohmic bath with an infinite cut-off frequency limit
$\Lambda \to \infty$. Such divergence causes covariances ($\Delta p, \Delta x$) of the Brownian particle to increase
immediately at the very beginning of the system evolution. In fact, all of Ohmic, sub-Ohmic and supra-Ohmic spectral
baths are suffered from the same divergence \cite{Zurek1989,Hu1992,Ford2001}.

In the literature \cite{Hu1992,Ford2001}, it has been argued that such initial jolt phenomenon is artificial and most
likely comes from the assumption of the initial decoupled system-environment state one used in deriving the master
equation. Here, we should prove that it is artificial but is not because of the use of the initial decoupled state. We
find that not only the diffusion coefficients in the generalized QBM, the decay rate also has similar behavior. While,
the decay rate is independent of the initial system-environment state. This can be seen from the combination of
Eqs.~(\ref{pGreenfs}), (\ref{renormph}) and (\ref{edcingeme}). It shows that the decay rate is determined purely by the
same dissipation dynamics described by the retarded nonequilibrium Green function Eq.~(\ref{uteq}) which is independent
of the initial system-environment state. This indicates that the problem of ``initial jolt'' has nothing to do with the
initial decoupled state that one often questioned \cite{Hu1992,Ford2001}. Below we will show how does the serious
divergence of the so-called ``initial jolt'' occurs and why it is not related to the choice of initial
system-environment states.

Plug in the Ohmic spectral density with an infinite cut-off into our exact master equation, the serious divergence
occurs in the decay coefficient $\gamma(t)$ and the fluctuation (diffusion) coefficients $\widetilde{\gamma}(t)$,
$\overline{\gamma}(t)$. To simplify the formulation, we take $J_{\scriptscriptstyle V}(\omega) \equiv 2 \pi \sum_k
\lvert V_k \rvert^2 \delta(\omega - \omega_k) = \sqrt{\frac{\pi\gamma_0}{2\Lambda}} \omega e^{-\omega/\Lambda}$ and
$\omega_\textsc{s} = \sqrt{2 \gamma_0 \Lambda / \pi}$. This setup corresponds to Ohmic spectral density with an
exponential cut-off and zero renormalized frequency in the conventional QBM model. The Fourier transformation of the
Ohmic spectral density is given by
\begin{align}
    g_{\scriptscriptstyle V}(t)
    \equiv
        \int \frac{\mathrm{d} \omega}{2 \pi} J_{\scriptscriptstyle V}(\omega) e^{- i \omega t}
    =
        \sqrt{\frac{\gamma_0}{8 \pi \Lambda}} (i t + 1/\Lambda)^{-2}
\end{align}
which has a pick in the time scale $1/\Lambda$. This function is related to the integral kernel in the time-convolution
equation for the generalized nonequilibrium Green function (\ref{pGreenfs}), which depicts back-action processes between
the system and the environment. The strength and the time scale of memory effect of the system is directly related to
the cut-off frequency $\Lambda$ which is the key to making the decay and diffusion coefficients diverge.

To show how this divergence is purely associated with the cut-off frequency introduced in the Ohmic spectral densities,
we evaluate the decay and diffusion coefficients $\gamma(t)$ and $\widetilde{\gamma}(t)$ by varying the ratio of
transition coupling strength and pairing strength $\lvert W_k / V_k \rvert = \alpha$.
Using Heisenberg equation of motion, one can easily find the equation of motion for the average occupation and the
squeezed parameter in arbitrary Brownian particle state,
\begin{align}
    & \frac{\mathrm{d}}{\mathrm{d} t}
    \begin{bmatrix}
        n(t) &
        s(t) \\
        s^*(t) &
        h(t)
    \end{bmatrix}
    \! =
        \biggl\{
            \!\! - i \omega_\textsc{s} \mathcal{Z}
            \begin{bmatrix}
                n(t) &
                s(t) \\
                s^*(t) &
                h(t)
            \end{bmatrix} \notag \\
            & \qquad\quad - i \mathcal{Z}
            \begin{bmatrix}
                V_k &
                W_k \\
                W_k^* &
                V_k^*
            \end{bmatrix}
            \begin{bmatrix}
                n_k'(t) &
                s_k'(t) \\
                s_k'^*(t) &
                n_k'^*(t)
            \end{bmatrix}
            + \text{H.c.}
        \biggr\}
    ,
\label{gcont2}
\end{align}
where $n_k'(t) = \langle a^\dagger(t) b_k(t) \rangle$ and $s_k'(t) = \langle a(t) b_k(t) \rangle$ are correlations
between the system and the environment. The first term represents the free evolution of the system, and the last term
represents the generalized information current flowing into the system. The later can be expressed explicitly as
\begin{align}
    \mathcal{I}(t)
    & \equiv
        \tfrac{\mathrm{d}}{\mathrm{d} t} \mathcal{N}(t,t)
        + \biggl\{
            i \omega_\textsc{s} \mathcal{Z} \mathcal{N}(t,t)
            + \text{H.c.}
        \biggr\} \notag \\
    & =
        \!\! \int_{t_0}^t \!\!\! \mathrm{d} \tau \Bigl\{
            \mathcal{Z} \widetilde{\mathcal{G}}(t,\tau) \mathcal{Z} \mathcal{U}^\dagger(t,\tau)
            - \mathcal{Z} \mathcal{G}(t,\tau) \mathcal{N}(\tau,t)
        \Bigr\}
        + \text{H.c.}
\label{discut}
\end{align}
after solving the environmental equation of motion. Here
$
    \mathcal{N}(\tau,t)
    \equiv
        \Bigl[ \begin{smallmatrix}
            \langle a^\dagger(t) a(\tau) \rangle &
            \langle a(t) a(\tau) \rangle \\
            \langle a^\dagger(t) a^\dagger(\tau) \rangle &
            \langle a(t) a^\dagger(\tau) \rangle
        \end{smallmatrix} \Bigr]
$
is the quadrature correlation matrix of the average occupation and the squeezed parameter. The first part in the
inegration is the diffusion current due to the initial distribution of the environment. The second part is the
dissipation current due to the initial state of the system, and the backflow from the system. Comparing
Eq.~(\ref{gcont}) with (\ref{discut}), the diffusion and decay coefficients $\widetilde{\gamma}(t)$,
$\overline{\gamma}(t)$ and $\gamma(t)$ can be also expressed as
\begin{align}
    \Bigl[
    \begin{matrix}
        \widetilde{\gamma}(t) &
        \overline{\gamma}(t) \\
        \overline{\gamma}^*(t) &
        \!\!\! \gamma(t) \!\! + \!\! \widetilde{\gamma}(t) \!
    \end{matrix}
    \Bigr]
    \!\! ={}
        & \!\!\!\! \int_{t_0}^t \!\!\!\! \mathrm{d} \tau \! \Bigl\{
            \! \mathcal{Z} \widetilde{\mathcal{G}}(t,\tau) \mathcal{U}(\tau,t) \mathcal{Z}
            \!\! - \!\! \mathcal{Z} \mathcal{G}(t,\tau) \mathcal{V}(\tau,t)
        \! \Bigr\} \notag \\
        & + \mathcal{K}(t) \mathcal{V}(t,t)
        + \text{H.c.}
\end{align}
At the beginning of the evolution, due to the initial condition $\mathcal{V}(t_0,t_0) = 0$ and $\mathcal{U}(t_0,t_0) =
I$, we only need to consider the first part in the integration, and the off-diagonal terms of $\mathcal{U}(\tau,t)$ can
be ignored. Then $\gamma(t)$ and $\widetilde{\gamma}(t)$ in the time scale $1/\Lambda$ can be approximated as
\begin{subequations}
\label{jolts}
\begin{align}
    & \widetilde{\gamma}(t)
    \approx
        2 \operatorname{Re} \left[
            \int_{t_0}^t \mathrm{d} \tau
            g_{\scriptscriptstyle W}(t,\tau)
            u_\textsc{s}(t,\tau)
        \right]
    ,
\label{jolts_gamt} \\
    & \gamma(t)
    \approx
        2 \operatorname{Re} \left[
            \int_{t_0}^t \mathrm{d} \tau
            (
                g_{\scriptscriptstyle V}^*(t,\tau) - g_{\scriptscriptstyle W}(t,\tau)
            )
            u_\textsc{s}(t,\tau)
        \right]
    ,
\label{jolts_gam}
\end{align}
\end{subequations}
where $u_\textsc{s}(t,\tau) \equiv \langle [a(t), a^\dagger(\tau)] \rangle = \mathcal{U}_{11}(t,\tau)$ and
$g_{\scriptscriptstyle W}(t,\tau) \equiv \sum_k \lvert W_k \rvert^2 e^{- i \omega_k (t - \tau)} = \alpha^2
g_{\scriptscriptstyle V}(t,\tau)$. Here we only discuss the low temperature case $k_\textsl{B} T / \hbar \ll \Lambda$,
so the terms proporsional to $\tilde{g}_{\scriptscriptstyle V}(t,\tau) \equiv \sum_k \lvert V_k \rvert^2
\overline{n}_k(t_0) e^{- i \omega_k (t - \tau)}$ and $\tilde{g}_{\scriptscriptstyle W}(t,\tau) \equiv \sum_k \lvert W_k
\rvert^2 \overline{n}_k(t_0) e^{- i \omega_k (t - \tau)}$ have also been dropped.

\begin{figure}
\subfigure[~$\gamma(t) / \Lambda$ in different $\alpha$]{
    \label{fig:jolts_gam}
    \includegraphics[width=0.45\linewidth]{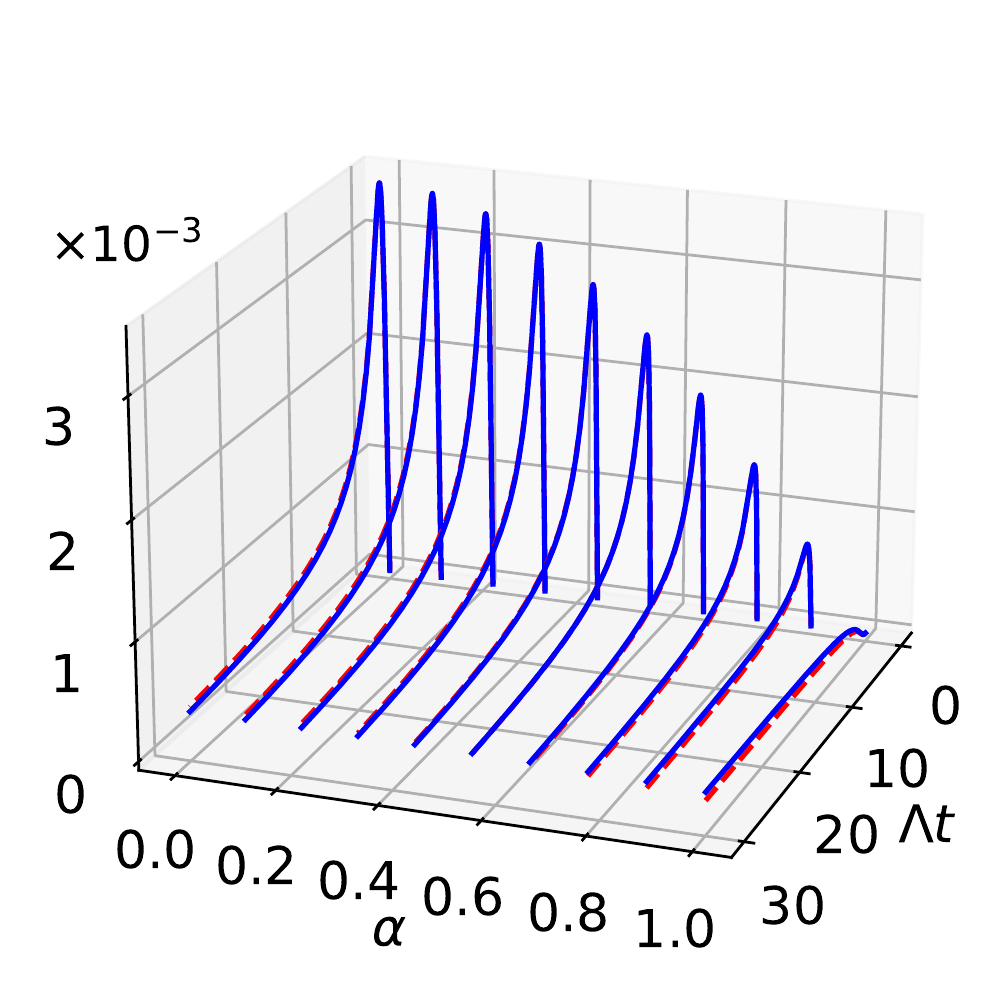}
}
\subfigure[~$\widetilde{\gamma}(t) / \Lambda$ in different $\alpha$]{
    \label{fig:jolts_gamt}
    \includegraphics[width=0.45\linewidth]{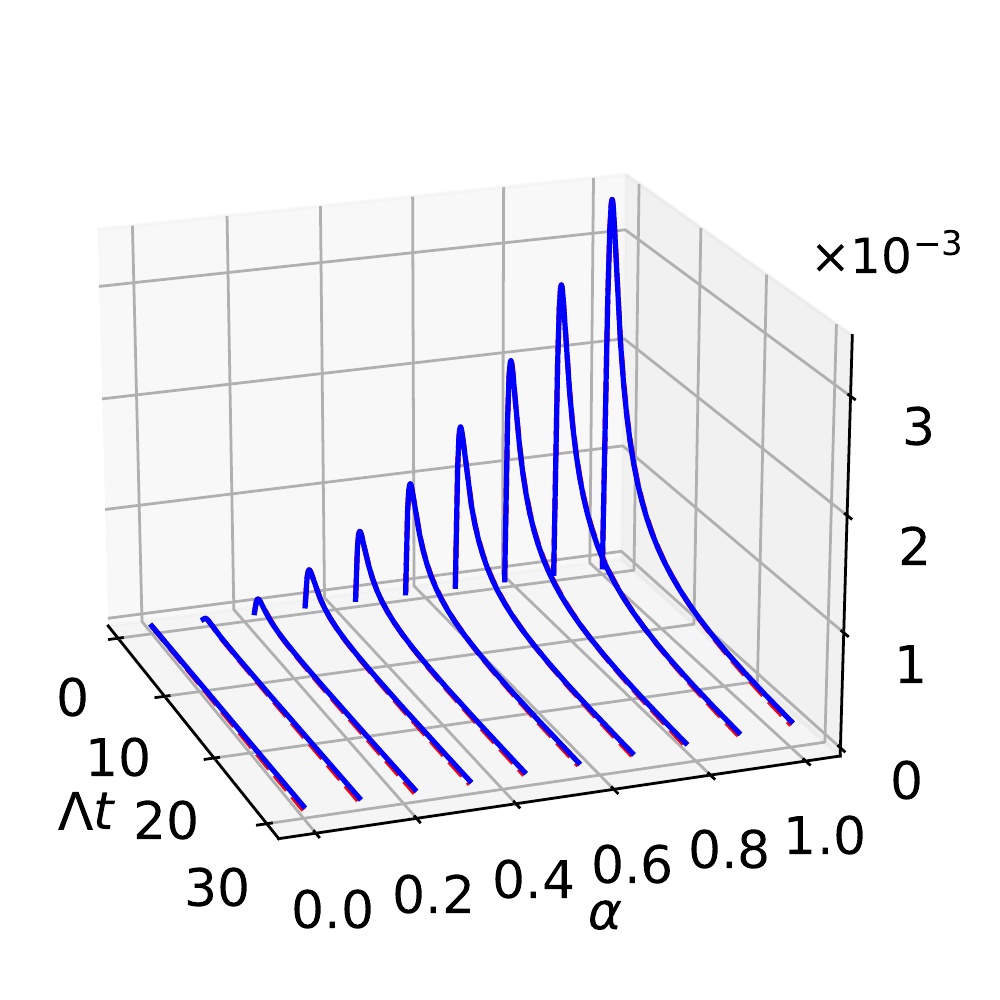}
}
\caption{
\label{fig:jolts}
(a) The decay coefficient $\gamma(t) / \Lambda$ and (b) the fluctuation coefficient $\widetilde{\gamma}(t) / \Lambda$ in
the time domain for different values of $\alpha = \lvert W_k / V_k \rvert$, the rate of the pairing strength with the
single particle coupling between the system and the environment in the generalized QBM. The exact solution (blue solid
line) from Eq.~(\ref{edcingeme}) and the estimated solution (red dashed line) based on Eq.~(\ref{jolts}) agree with each
other. Here we set $\gamma_0 = 0.0003 \Lambda$, $k_\textsl{B} T / \hbar = 0.01 \Lambda$. They are plotted against time
scale $1 / \Lambda$ to emphasize the initial jolt. The results show that the initial jolt exists in both the decay
coefficient $\gamma(t)$ and the diffusion coefficient $\widetilde{\gamma}(t)$.
}
\end{figure}

We show numerically that the estimated results of the diffusion and decay coefficients $\widetilde{\gamma}(t)$ and
$\gamma(t)$ calculated from Eq.~(\ref{jolts}) are almost the same as the exact solutions of Eq.~(\ref{edcingeme}) for
all the values of $\alpha$: $0 \leq \alpha \leq 1$ at low temperature with a high cut-off frequency, see
Fig.~\ref{fig:jolts}.
When $\alpha = 1$, the amplitudes of $g_{\scriptscriptstyle V}(t,\tau)$ and $g_{\scriptscriptstyle W}(t,\tau)$ become
the same, then the divergences in the decay coefficient is cancelled, as analytically shown by Eq.~(\ref{jolts_gam}).
Figure \ref{fig:jolts_gam} clearly show that for $\alpha = 1$, the decay coefficient $\gamma(t)$ suddenly increases only
a little bit within the time scale $1/\Lambda$ but then it approaches to $\gamma_0$ instead of decreasing. In other
words, no ``initial jolt'' occurs in the dissipation $\gamma(t)$ for the conventional QBM because of the accident
cancellation of the divergences. For any other $\alpha$ values (which is more realistic in quantum optics), the
divergence cannot be fully cancelled so that the decay coefficient $\gamma(t)$ also shows the ``initial jolt'', as given
in Fig.~\ref{fig:jolts_gam}. Because the decay coefficient $\gamma(t)$ is independent of the initial system-environment
state. We conclude that the ``initial jolt'' has nothing to do with the initial decoupled state one used in deriving the
exact master equation. 

With the finding of this accident cancellation of the divergences in the decay coefficient $\gamma(t)$ due to the
special choice of $\alpha = 1$ in the QBM, the origin of the initial jolt becomes clear. It comes purely from the
Ohmic-type spectral densities with the infinite cut-off frequency $\Lambda \rightarrow \infty$. For a fundamental
renormalizable theory, one requires that any renormalized quantity must be independent of regularization-scheme, so that
the high energy cut-off scale can be taken to the infinite limit in the calculations of energy. However, both the
conventional and the generalized QBM are build on phenomenological model Hamiltonians, no such renormalizability is
required in priori. Furthermore, the cut-off frequency in the Ohmic-type spectral densities is introduced to count the
effective transition processes between the system and the environment. Quantum mechanically, the most active
environmental modes $\omega_k$ that effectively couple to the system are dominated by these not far away from the
thermal frequency $k_\textsl{B} T / \hbar$ so that physical transition between the system and the environment can occur
(with large probabilities). Physically, the spectral density is a summation of the probabilities over the physical
processes between the system and the environment. For the Ohmic-type spectral densities, taking a very large cut-off
frequency implies that the environmental modes with higher energy will have the larger probability amplitude for
coupling to the system mode. This is obviously very unphysical. The physical reliable cut-off frequency should be
estimated by the condition that: the mean frequency of a given spectral density should be the same order of the system
characterized frequency, if the system-environment coupling cannot be derived from a more fundamental theory or no
experimental data can be used to fix it. Under such a physical requirement, the problem of ``initial jolt'' will never
occur.

\section{Discussions and Perspectives}
\label{sec:conclusion}
In this paper, we provide a detailed analysis on system-environmental couplings which are usually unknown in priori but
are the key to the determination of dissipation and fluctuation dynamics in open quantum systems. In particular, we show
that the quantum dissipative dynamics described by Feynman-Vernon model and Caldeira-Leggett model based on classical
mechanics involve some physically inconsistent processes at quantum level. These inconsistent processes include the
transitions from low-energy states to high-energy states by emitting an amount of energy or create particles from
nothing, which should be forbidden in quantum mechanics. Within the framework of quantum mechanics, each individual
microscopic particle process should be constrained by the conservations of energy-momentum and angular momentum, etc. We
further show that one can get rid of these inconsistent processes by including momentum-dependent system-environment
couplings. We generalize the QBM to include such momentum-dependent couplings, which are indeed easier to be realized in
realistic quantum world. We derive the exact master equation of such generalized QBM for both the initially decoupled
and initially correlated system-environment states.

With the generalized QBM and its exact master equation for both the initially decoupled and the initially correlated
state, we reproduce the Hu-Paz-Zhang master equation as a special case. We then find that in the Hu-Paz-Zhang equation
for the conventional QBM with the position-dependent coupling only, the renormalized Brownian particle Hamiltonian
actually contains a renormalization-induced momentum-dependent potential. This momentum-dependent potential is misplaced
into the dissipation term in the Hu-Paz-Zhang equation, so that the correct renormalized Brownian particle Hamiltonian
was not found before. The environment induced momentum-dependent potential in the Hu-Paz-Zhang equation also indicates
that including the momentum-dependent system-environment coupling is a natural consequence of the QBM.

Furthermore, we re-examine the long-standing problem of the initial jolt in the conventional QBM and Hu-Paz-Zhang master
equation. The initial jolt is related to the divergence arising from the Ohmic-type spectral density with an infinite
cut-off frequency. It has been thought that the initial jolt is a result of using the initial decoupled state which is
physically unacceptable, but we find that this is indeed a misunderstanding. The misunderstanding comes from the
accident cancellation of the divergences in the decay coefficient in the Hu-Paz-Zhang equation, due to the same coupling
strengths for the single particle transition and the pairing processes. With the generalized QBM including the
momentum-dependent coupling, we show that the initial jolt exists in both the decay and diffusion coefficients, while
the decay coefficient is independent of the initial system-environment state. As a conclusion, the so-called ``initial
jolt'' has nothing do to with the initial decoupled state. It is an artificial effect when one takes an extremely large
cut-off frequency in comparison with the thermal energy of the environment, which is unphysical as we explained in the
end of the last section.

With the generalized QBM, we have resolved these interesting and important problems for quantum dissipative dynamics, as
we discussed in this work. The new exact master equation for the generalized QBM also has the potential applications to
photonics quantum computing. With an extension of the generalized QBM to multimodes, one can apply it to the integrated
quantum processor made of many squeezers and interferometers with hundred more photonic modes, that currently attracted
a great attention in the development of quantum technology. We will leave this problem for further investigation.


\acknowledgments

This work is supported by Ministry of Science and Technology of Taiwan, Republic of China under Contract No.
MOST-108-2112-M-006-009-MY3.

\appendix
\section{The generalized nonequilibrium Green functions for open quantum systems with pairing couplings and
fluctuation-dissipation theorem}
\label{sec:UV}
The nonequilibrium Green functions we introduced for open quantum systems \cite{Tu2008,Jin2010,Lei2012,Zhang2012} are
generalized to open quantum systems involving pairing couplings in this work, and are denoted by $\mathcal{U}(\tau,t_0)$
and $\mathcal{V}(\tau,t)$ \cite{Huang2020}. They are defined as
\begin{align}
    & \mathcal{U}(t,t_0) \mathcal{Z}
    =
        \begin{bmatrix}
            \langle [a(t), a^\dagger(t_0)] \rangle &
            \langle [a(t), a(t_0)] \rangle \\
            \langle [a^\dagger(t), a^\dagger(t_0)] \rangle &
            \langle [a^\dagger(t), a(t_0)] \rangle
        \end{bmatrix}
    ,
\label{udef} \\
    & \mathcal{V}(\tau,t)
    =
        \begin{bmatrix}
            \langle a^\dagger(t) a(\tau) \rangle &
            \langle a(t) a(\tau) \rangle \\
            \langle a^\dagger(t) a^\dagger(\tau) \rangle &
            \langle a(t) a^\dagger(\tau) \rangle
        \end{bmatrix} \notag \\
        & \qquad\qquad\qquad\qquad - (\text{initial-dependent part})
    ,
\end{align}
and can be easily determined directly from Heisenberg equation of motion for the generalized QBM Hamiltomnian
Eq.~(\ref{mCLH}),
\begin{subequations}
\begin{align}
    & \frac{\mathrm{d}}{\mathrm{d} t}
    \! \begin{bmatrix}
        a(t) \\
        a^\dagger(t)
    \end{bmatrix}
    =
        - i \omega_\textsc{s} \mathcal{Z}
        \! \begin{bmatrix}
            a(t) \\
            a^\dagger(t)
        \end{bmatrix}
        \! - \! i \sum_k \mathcal{Z}
        \! \begin{bmatrix}
            V_k & W_k \\
            W_k^* & V_k^* \\
        \end{bmatrix}
        \! \begin{bmatrix}
            b_k(t) \\
            b_k^\dagger(t)
        \end{bmatrix}
    ,
\label{aeq} \\
    & \frac{\mathrm{d}}{\mathrm{d} t}
    \! \begin{bmatrix}
        b_k(t) \\
        b_k^\dagger(t)
    \end{bmatrix}
    =
        - i \omega_k \mathcal{Z}
        \! \begin{bmatrix}
            b_k(t) \\
            b_k^\dagger(t)
        \end{bmatrix}
        \! - \! i \mathcal{Z}
        \! \begin{bmatrix}
            V_k^* & W_k \\
            W_k^* & V_k \\
        \end{bmatrix}
        \! \begin{bmatrix}
            a(t) \\
            a^\dagger(t)
        \end{bmatrix}
    .
\label{bkeq}
\end{align}
\end{subequations}
These equations of motion determine all the dissipation and fluctuation dynamics of the Brownian particle.

Formally, solving Eq.~(\ref{bkeq}), we obtain
\begin{align}
    \begin{bmatrix}
        b_k(t) \\
        b_k^\dagger(t)
    \end{bmatrix}
    & =
        \begin{bmatrix}
            e^{- i \omega_k (t - t_0)} & \!\! 0 \\
            0 & \!\! e^{+ i \omega_k (t - t_0)}
        \end{bmatrix}
        \! \begin{bmatrix}
            b_k(t_0) \\
            b_k^\dagger(t_0)
        \end{bmatrix} \notag \\
        & \! - \! i \!\! \int_{t_0}^t \!\! \mathrm{d} \tau
        \mathcal{Z} \! \begin{bmatrix}
            V_k^*(\tau-t) & W_k(\tau-t) \\
            W_k^*(\tau-t) & V_k(\tau-t) \\
        \end{bmatrix}
        \! \begin{bmatrix}
            a(\tau) \\
            a^\dagger(\tau)
        \end{bmatrix}
    .
\end{align}
This formal solution allows us to eliminate exactly and completely all the environmental degrees of freedom (equivalent
to the trace over all the environment states). Substituting this solution into Eq.~(\ref{aeq}), we have
\begin{align}
    & \frac{\mathrm{d}}{\mathrm{d} t}
    \begin{bmatrix}
        a(t) \\
        a^\dagger(t)
    \end{bmatrix}
    + i \omega_\textsc{s} \mathcal{Z}
    \begin{bmatrix}
        a(t) \\
        a^\dagger(t)
    \end{bmatrix}
    \! + \!\! \int_{t_0}^t \!\! \mathrm{d} \tau \mathcal{Z}
    \mathcal{G}(t,\tau)
    \! \begin{bmatrix}
        a(\tau) \\
        a^\dagger(\tau)
    \end{bmatrix} \notag \\
    & \qquad\qquad =
        - i \sum_k \mathcal{Z}
        \! \begin{bmatrix}
            V_k(t-t_0) & W_k(t-t_0) \\
            W_k^*(t-t_0) & V_k^*(t-t_0) \\
        \end{bmatrix}
        \! \begin{bmatrix}
            b_k(t_0) \\
            b_k^\dagger(t_0)
        \end{bmatrix}
    ,
\label{aaeq}
\end{align}
where
\begin{align}
    \mathcal{G}(\tau,\tau')
    & =
        \!\! \sum_k
        \! \begin{bmatrix}
            V_k(\tau\!-\!t_0) & \!\! W_k(\tau\!-\!t_0) \\
            W_k^*(\tau\!-\!t_0) & \!\! V_k^*(\tau\!-\!t_0)
        \end{bmatrix} \notag \\
        & \qquad \times \!\! \begin{bmatrix} 1 & 0 \\ 0 & -1 \end{bmatrix}
        \!\! \begin{bmatrix}
            V_k^*(\tau'\!-\!t_0) & \!\! W_k(\tau'\!-\!t_0) \\
            W_k^*(\tau'\!-\!t_0) & \!\! V_k(\tau'\!-\!t_0)
        \end{bmatrix}
\label{gqle}
\end{align}
is Eq.~(\ref{setc}). One can clearly see that Eq.~(\ref{aaeq}) is the generalized quantum Langevin equation, in which
the third term in the left hand side is the dissipation (decay) term due to the system-environment coupling, and the
term in the right hand side is the fundamental noise force generated by the environment through the system-environment
coupling. It is obvious that the average of the noise force is zero if the environment is initially in a thermal state.

Because Eq.(\ref{aaeq}) is a linear differential equation, its general solution has the form
\begin{align}
    \begin{bmatrix}
        a(t) \\
        a^\dagger(t)
    \end{bmatrix}
    =
        \mathcal{U}(t,t_0)
        \begin{bmatrix}
            a(t_0) \\
            a^\dagger(t_0)
        \end{bmatrix}
        + \begin{bmatrix}
            F(t) \\
            F^\dagger(t)
        \end{bmatrix}
    .
\label{aags}
\end{align}
Substituting this general solution into Eq.~(\ref{aaeq}), one can find that the retarded Green function
$\mathcal{U}(t,t_0)$ describes the dissipation dynamics of the Brownian particle and obeys the Dyson equation of motion,
\begin{align}
    \frac{\mathrm{d}}{\mathrm{d} \tau} \mathcal{U}(\tau,t_0)
    \! + \! i \omega_\textsc{s} \mathcal{Z} \mathcal{U}(\tau,t_0)
    \! + \!\!\int_{t_0}^\tau \!\!\! \mathrm{d} \tau' \mathcal{Z} \mathcal{G}(\tau,\tau') \mathcal{U}(\tau',t_0)
    = 0
    ,
\end{align}
which is just Eq.~(\ref{uteq}). The integral kernel in the above equation characterize the non-Markovian dissipation
processes. It is easy to show that the definition of the retarded Green function $\mathcal{U}(t,t_0)$ in
Eq.~(\ref{aags}) is the same as the definition of Eq.~(\ref{udef}),
\begin{align}
    & \begin{bmatrix}
        \langle [a(t), a^\dagger(t_0)] \rangle &
        \langle [a(t), a(t_0)] \rangle \\
        \langle [a^\dagger(t), a^\dagger(t_0)] \rangle &
        \langle [a^\dagger(t), a(t_0)] \rangle
    \end{bmatrix} \notag \\
    & \qquad =
        \mathcal{U}(t,t_0)
        \begin{bmatrix}
            \langle [a(t_0), a^\dagger(t_0)] \rangle &
            \langle [a(t_0), a(t_0)] \rangle \\
            \langle [a^\dagger(t_0), a^\dagger(t_0)] \rangle &
            \langle [a^\dagger(t_0), a(t_0)] \rangle
        \end{bmatrix} \notag \\
    & \qquad =
        \mathcal{U}(t,t_0) \mathcal{Z}
    .
\end{align}

The last term in Eq.~(\ref{aags}) describes the environment-induced fluctuation (noise) dynamics that obeys the
following equation of motion,
\begin{align}
    & \frac{\mathrm{d}}{\mathrm{d} t}
    \begin{bmatrix}
        F(t) \\
        F^\dagger(t)
    \end{bmatrix}
    + i \omega_\textsc{s} \mathcal{Z}
    \begin{bmatrix}
        F(t) \\
        F^\dagger(t)
    \end{bmatrix}
    \! + \!\! \int_{t_0}^t \!\! \mathrm{d} \tau
    \mathcal{Z} \mathcal{G}(t,\tau)
    \! \begin{bmatrix}
        F(\tau) \\
        F^\dagger(\tau)
    \end{bmatrix} \notag \\
    & \qquad\qquad =
        - i \sum_k \mathcal{Z}
        \begin{bmatrix}
            V_k(t-t_0) & W_k(t-t_0) \\
            W_k^*(t-t_0) & V_k^*(t-t_0) \\
        \end{bmatrix}
        \! \begin{bmatrix}
            b_k(t_0) \\
            b_k^\dagger(t_0)
        \end{bmatrix}
    ,
\end{align}
subjected to the initial condition $\Big[ \begin{smallmatrix} F(t_0) \\ F^\dagger(t_0) \end{smallmatrix} \Big] = 0$.
From this initial condition, the general solution of the equation for the noise dynamics is given explicitly,
\begin{align}
    & \begin{bmatrix}
        F(t) \\
        F^\dagger(t)
    \end{bmatrix} \notag \\
    & =
        - i \! \sum_k \!\! \int_{t_0}^t \!\! \mathrm{d} \tau
        \mathcal{U}(t,\tau) \mathcal{Z}
        \! \begin{bmatrix}
            V_k(t-t_0) & \! W_k(t-t_0) \\
            W_k^*(t-t_0) & \! V_k^*(t-t_0) \\
        \end{bmatrix}
        \!\! \begin{bmatrix}
            b_k(t_0) \\
            b_k^\dagger(t_0)
        \end{bmatrix}
    .
\end{align}
The nonequilibrium correlation functions are defined by
\begin{align}
    & \begin{bmatrix}
        \langle a^\dagger(t) a(\tau) \rangle &
        \langle a(t) a(\tau) \rangle \\
        \langle a^\dagger(t) a^\dagger(\tau) \rangle &
        \langle a(t) a^\dagger(\tau) \rangle
    \end{bmatrix} \notag \\
    & \qquad =
        \mathcal{U}(\tau,t_0)
        \begin{bmatrix}
            \langle a^\dagger(t_0) a(t_0) \rangle &
            \langle a(t_0) a(t_0) \rangle \\
            \langle a^\dagger(t_0) a^\dagger(t_0) \rangle &
            \langle a(t_0) a^\dagger(t_0) \rangle
        \end{bmatrix}
        \mathcal{U}^\dagger(t,t_0)
        \notag \\
        & \qquad\qquad\qquad + \begin{bmatrix}
            \langle F^\dagger(t) F(\tau) \rangle &
            \langle F(t) F(\tau) \rangle \\
            \langle F^\dagger(t) F^\dagger(\tau) \rangle &
            \langle F(t) F^\dagger(\tau) \rangle
        \end{bmatrix}
    .
\end{align}
The last term is the noise-induced correlation Green function $\mathcal{V}(\tau,t)$, which has the solution
\begin{align}
    \mathcal{V}(\tau,t)
    & =
        \begin{bmatrix}
            \langle F^\dagger(t) F(\tau) \rangle &
            \langle F(t) F(\tau) \rangle \\
            \langle F^\dagger(t) F^\dagger(\tau) \rangle &
            \langle F(t) F^\dagger(\tau) \rangle
        \end{bmatrix} \notag \\
    & =
        \!\! \int_{t_0}^\tau \!\! \mathrm{d}\tau'
        \!\! \int_{t_0}^t \!\! \mathrm{d} t'
        \mathcal{U}(\tau,\tau') \mathcal{Z} \widetilde{\mathcal{G}}(\tau',t') \mathcal{Z} \mathcal{U}^\dagger(t,t')
    .
\label{vtss}
\end{align}
This gives the same solution Eq.~(\ref{vtsol}) obtained in our exact master equation derivation. The
initial-state-dependent system-environment correlation function $\widetilde{\mathcal{G}}(t,t')$ is given by
\begin{align}
    \widetilde{\mathcal{G}}(\tau',t')
    & =
        \sum_k \begin{bmatrix}
            V_k(\tau'-t_0) & W_k(\tau'-t_0) \\
            W_k^*(\tau'-t_0) & V_k^*(\tau'-t_0) \\
        \end{bmatrix} \notag \\
        & \qquad\quad \times
        \begin{bmatrix}
            \langle b_k^\dagger(t) b_k(\tau) \rangle &
            \langle b_k(t) b_k(\tau) \rangle \\
            \langle b_k^\dagger(t) b_k^\dagger(\tau) \rangle &
            \langle b_k(t) b_k^\dagger(\tau) \rangle
        \end{bmatrix} \notag \\
        & \qquad\quad \times \begin{bmatrix}
            V_k^*(t'-t_0) & W_k(t'-t_0) \\
            W_k^*(t'-t_0) & V_k(t'-t_0) \\
        \end{bmatrix}
    ,
\end{align}
which is Eq.~(\ref{idsetc}). The solution Eq.~(\ref{vtss}) is also the generalization of the Keldysh's correlation
function in the nonequilibrium technique \cite{Keldysh1965}. It is indeed the generalized nonequilibrium
fluctuation-dissipation theorem, as a consequence of the unitarity principle of quantum mechanics for the whole system
(the system plus the environment) \cite{Zhang2012}. The usual equilibrium fluctuation-dissipation theorem can be easily
obtained from this solution as a steady-state limit.

\end{document}